%% file: main.tex
\documentclass[acmsmall,screen,nonacm]{acmart}

\usepackage{quantikz}
\definecolor{qubitblue}{rgb}{0.0, 0.0, 0.55}
\definecolor{maroon}{rgb}{0.5, 0.0, 0.0}
\newcommand{\q}[1]{{\color{qubitblue}{#1}}}
\usepackage{algorithm}
\usepackage[noend]{algpseudocode}
\usepackage[capitalise,nameinlink]{cleveref}
\crefname{figure}{Fig.}{Figs.}
\Crefname{figure}{Fig.}{Figs.}
\crefname{table}{Tbl.}{Tbls.}
\Crefname{table}{Tbl.}{Tbls.}
\crefname{corollary}{Corollary}{Corollaries}
\Crefname{corollary}{Corollary}{Corollaries}
\crefname{lemma}{Lemma}{Lemmas}
\Crefname{lemma}{Lemma}{Lemmas}
\crefname{section}{Sec.}{Secs.}
\Crefname{section}{Sec.}{Secs.}
\crefname{subsection}{Sec.}{Secs.}
\Crefname{subsection}{Sec.}{Secs.}
\crefname{subsubsection}{Sec.}{Secs.}
\Crefname{subsubsection}{Sec.}{Secs.}
\crefname{appendix}{App.}{Apps.}
\Crefname{appendix}{App.}{Apps.}
\crefname{subappendix}{App.}{Apps.}
\Crefname{subappendix}{App.}{Apps.}
\crefname{subsubappendix}{App.}{Apps.}
\Crefname{subsubappendix}{App.}{Apps.}
\AddToHook{env/definition/begin}{\crefalias{theorem}{definition}}
\AddToHook{env/example/begin}{\crefalias{theorem}{example}}
\usepackage{xspace}
\usepackage{array}
\usepackage{subcaption}
\usepackage{tikz}
\usepackage{wrapfig}
\usepackage{stmaryrd}
\usepackage{mathpartir}
\usetikzlibrary{arrows.meta}
\usepackage{listings}
\definecolor{leankw}{HTML}{534AB7}\definecolor{leantype}{HTML}{0F6E56}
\definecolor{leantactic}{HTML}{B4540A}
\lstdefinelanguage{lean}{
  morekeywords=[1]{def,structure,inductive,where,extends,deriving,instance,theorem,
    lemma,noncomputable,abbrev,namespace,end,variable,fun,let,by,open},
  morekeywords=[2]{Type,Prop,Sort,Fin,Finset,List,Set,Nat,Subgroup},
  morekeywords=[3]{decide,rfl,simp,exact,omega,intro,rw,apply,refine},
  morecomment=[l]{--},
  morecomment=[s]{/-}{-/},
  sensitive=true,
  literate={→}{{$\to$}}1 {ℕ}{{\textcolor{leantype}{$\mathbb{N}$}}}1 {⟨}{{$\langle$}}1 {⟩}{{$\rangle$}}1
    {∀}{{$\forall$}}1 {∈}{{$\in$}}1 {∉}{{$\notin$}}1 {≤}{{$\le$}}1
    {σ}{{$\sigma$}}1 {…}{{$\ldots$}}1 {≠}{{$\neq$}}1 {≥}{{$\geq$}}1
    {∧}{{$\wedge$}}1 {∃}{{$\exists$}}1 {↦}{{$\mapsto$}}1
    {∂}{{$\partial$}}1 {₁}{{$_1$}}1 {₂}{{$_2$}}1 {≃}{{$\simeq$}}1 {ₗ}{{$_{\ell}$}}1 {∪}{{$\cup$}}1
    {ℓ}{{$\ell$}}1 {×}{{$\times$}}1 {⊓}{{$\sqcap$}}1
    {:=}{{\raisebox{0.2ex}{:}\kern-0.15em=}}2,
}
\makeatletter
\lst@AddToHook{OnEmptyLine}{\vspace{-0.5\baselineskip}}
\makeatother
\usepackage{tcolorbox}\tcbuselibrary{listings,skins}
\newtcblisting{leancode}{listing only, blanker,
  borderline west={1.5pt}{0pt}{black!25}, left skip=0.9em, left=0.5em+6pt,
  before skip=0.6em, after skip=0.6em,
  listing options={language=lean, aboveskip=0pt, belowskip=0pt}}
\definecolor{specfill}{HTML}{EEEDFE}\definecolor{specdraw}{HTML}{534AB7}\definecolor{spectext}{HTML}{3C3489}
\definecolor{viewfill}{HTML}{E1F5EE}\definecolor{viewdraw}{HTML}{0F6E56}\definecolor{viewtext}{HTML}{085041}
\definecolor{instfill}{HTML}{F1EFE8}\definecolor{instdraw}{HTML}{5F5E5A}\definecolor{insttext}{HTML}{444441}
\definecolor{xstab}{HTML}{F28B82}\definecolor{zstab}{HTML}{8AB4F8}
\tikzset{chk/.style={draw=black!70, line width=0.3pt, fill opacity=0.55}}
\DeclareMathOperator{\im}{img}
\newcommand{\surfacepatch}[1]{%
  \pgfmathtruncatemacro{\dm}{#1-1}%
  \pgfmathtruncatemacro{\dmm}{#1-2}%
  \foreach \i in {0,...,\dmm} {%
    \pgfmathtruncatemacro{\p}{mod(\i+\dm,2)}%
    \ifnum\p=0
      \filldraw[chk, fill=xstab] (\i,\dm) arc[start angle=180, end angle=0, radius=0.5] -- cycle;
    \fi
    \pgfmathtruncatemacro{\p}{mod(\i,2)}%
    \ifnum\p=1
      \filldraw[chk, fill=xstab] (\i,0) arc[start angle=180, end angle=360, radius=0.5] -- cycle;
    \fi
  }%
  \foreach \j in {0,...,\dmm} {%
    \pgfmathtruncatemacro{\p}{mod(\j,2)}%
    \ifnum\p=0
      \filldraw[chk, fill=zstab] (0,\j) arc[start angle=-90, end angle=-270, radius=0.5] -- cycle;
    \fi
    \pgfmathtruncatemacro{\p}{mod(\j+\dm,2)}%
    \ifnum\p=1
      \filldraw[chk, fill=zstab] (\dm,\j) arc[start angle=-90, end angle=90, radius=0.5] -- cycle;
    \fi
  }%
  \foreach \i in {0,...,\dmm} \foreach \j in {0,...,\dmm} {%
    \pgfmathtruncatemacro{\p}{mod(\i+\j,2)}%
    \ifnum\p=0 \def\c{xstab}\else\def\c{zstab}\fi
    \filldraw[chk, fill=\c] (\i,\j) rectangle ++(1,1);
  }%
  \foreach \i in {0,...,\dm} \foreach \j in {0,...,\dm}
    \fill[black!85] (\i,\j) circle[radius=0.12];
}
\definecolor{ccred}{HTML}{F28B82}\definecolor{ccgreen}{HTML}{81C995}\definecolor{ccblue}{HTML}{8AB4F8}
\newcommand{\colorcodepatch}[1]{%
  \pgfmathtruncatemacro{\side}{3*(#1-1)/2}%
  \begin{scope}[x={(1,0)}, y={(0.5,0.8660)}]
    \begin{scope}
      \clip (0,0) -- (\side,0) -- (0,\side) -- cycle;
      \foreach \a in {0,...,\side} \foreach \b in {0,...,\side} {%
        \pgfmathtruncatemacro{\p}{mod(\a+2*\b,3)}%
        \pgfmathtruncatemacro{\inside}{\a+\b<=\side}%
        \ifnum\p=1 \ifnum\inside=1
          \pgfmathtruncatemacro{\k}{mod(\a,3)}%
          \ifcase\k \def\fc{ccred}\or\def\fc{ccgreen}\or\def\fc{ccblue}\fi
          \filldraw[chk, fill=\fc] (\a+1,\b) -- (\a,\b+1) -- (\a-1,\b+1)
            -- (\a-1,\b) -- (\a,\b-1) -- (\a+1,\b-1) -- cycle;
        \fi\fi
      }%
    \end{scope}
    \draw[chk] (0,0) -- (\side,0) -- (0,\side) -- cycle;
    \foreach \a in {0,...,\side} \foreach \b in {0,...,\side} {%
      \pgfmathtruncatemacro{\p}{mod(\a+2*\b,3)}%
      \pgfmathtruncatemacro{\inside}{\a+\b<=\side}%
      \ifnum\p=1 \else \ifnum\inside=1
        \fill[black!85] (\a,\b) circle[radius=1.6pt];
      \fi\fi
    }%
  \end{scope}
}
\newcommand{\toricgrid}[4]{%
  \pgfmathtruncatemacro{\lm}{#1-1}%
  \foreach \k in {0,...,\lm}
    \draw[black!70, line width=0.4pt] (\k,-0.5) -- (\k,\lm+0.5) (-0.5,\k) -- (\lm+0.5,\k);
  \draw[ccred, line width=1.6pt, line cap=round] (#2,0) rectangle ++(1,1);
  \begin{scope}[ccblue, line width=1.6pt, line cap=round]
    \draw (#3,#4-1) -- (#3,#4+1) (#3,#4) -- (#3+1,#4);
    \ifnum#3=0
      \draw (-0.5,#4) -- (0,#4) (\lm,#4) -- (\lm+0.5,#4);
    \else
      \draw (#3-1,#4) -- (#3,#4);
    \fi
  \end{scope}
}

\newcommand{\f}{QECLean\xspace}
\newcommand{\nkd}[1]{\llbracket #1 \rrbracket}
\newcommand{\Cbase}{C_{\mathrm{base}}}
\newcommand{\Ccov}{C_{\mathrm{cover}}}
\DeclareRobustCommand{\ruleConcat}{\hyperlink{rule:concat}{\textsc{Concat}}\xspace}
\DeclareRobustCommand{\ruleDouble}{\hyperlink{rule:double}{\textsc{Double}}\xspace}
\DeclareRobustCommand{\ruleCover}{\hyperlink{rule:cover}{\textsc{Cover}}\xspace}

\begin{document}
\title{Specifying \& Verifying Quantum Error-Correcting Codes}

\author{Stavan Jain}
\affiliation{%
  \institution{University of Wisconsin--Madison}
  \city{Madison}
  \state{WI}
  \country{USA}}
\email{sjain298@wisc.edu}
\author{Nithin Raveendran}
\affiliation{%
  \institution{University of Arizona}
  \city{Tucson}
  \state{AZ}
  \country{USA}}
\email{nithin@arizona.edu}
\author{Aws Albarghouthi}
\affiliation{%
  \institution{University of Wisconsin--Madison}
  \city{Madison}
  \state{WI}
  \country{USA}}
\email{aws@cs.wisc.edu}

\begin{abstract}
Building quantum computers is challenging due to the noise that affects qubits.
Quantum error correction tackles noise by encoding information in quantum error-correcting (\textsc{qec}) codes.
These codes are often complex, increasingly found by computational search, presented in diverse formalisms with informal arguments, and hard to reason about algorithmically.
Given that the rest of the quantum computing stack relies on a \textsc{qec} code's properties for correct computation, we propose a formally verified foundation for \textsc{qec} codes.

We present \f, a Lean-based framework for formally specifying and verifying \textsc{qec} codes.
\f backs every claimed \textsc{qec}-code property with a machine-checked proof, unifies the diverse presentations of codes in the literature through common abstractions, and makes new codes easy to specify.
We present a new compositional technique, \emph{code combinators}, for proving code distance analytically.
Code combinators allow us to prove the distance of large codes that are out of reach for algorithmic techniques; for example, we show that \textsc{ibm}'s $\nkd{360, 12, \leq 24}$ code has distance  24, which was previously only known as an upper bound. Furthermore, we show how our approach allows us to formalize and prove entire code families, such as the surface code.
Our work lays the foundation required for reasoning about circuit-level implementations of codes on given architectures and for building end-to-end verified compilers for quantum computers.
\f is available at \url{https://github.com/Stavan-Jain/QECLean}.
\end{abstract}
\maketitle

\section{Introduction}
Quantum computing promises asymptotic speedups for important problems in domains like quantum chemistry and cryptography~\citep{lloyd1996universal,shor1994algorithms}.
Practically realizing these speedups requires addressing the high physical error rates of quantum computers.
The field of quantum error correction~\citep{shor1995scheme,gottesman1997stabilizer} provides us with recipes for detecting and correcting errors in quantum computers.

The key algorithmic problem in building error-corrected quantum computers is designing a \emph{quantum error-correcting} (\textsc{qec}) code, which, like its classical counterparts, encodes logical information by introducing redundancy at the physical level.
There are three key properties that dictate how useful a \textsc{qec} code is:
(1) \emph{Hardware compatibility}: The \textsc{qec} code has to satisfy the constraints imposed by the underlying hardware.
(2) \emph{Encoding rate}: The \textsc{qec} code has to encode as much logical information in as few physical qubits as possible.
(3) \emph{Distance}: The \textsc{qec} code has to have a high \emph{code distance}, which is, roughly speaking, a measure of how many errors a \textsc{qec} code can withstand. The latter two properties are often described in the \textsc{qec} literature by the $\nkd{n, k, d}$ notation~\citep{calderbank1998quantum}: a \textsc{qec} code is an $\nkd{n, k, d}$ code if it encodes $k$ logical qubits, using $n$ physical qubits, with code distance $d$.

The parameters of a \textsc{qec} code are a guarantee the rest of the quantum computing stack relies on: hardware architectures are built to implement the code's measurements, and compilation assumes its distance to establish sufficiently low error rates.
Yet \textsc{qec} code parameters are hard to establish.
Computing the distance of a code is \textsc{np}-hard in general~\citep{kapshikar2023hardness}, so the distance of a large code is often reported only as an upper bound, obtained by simulation or brute-force search.
Unfortunately, such a bound says nothing about how many errors the code actually withstands.
For example, one of \textsc{ibm}'s \emph{bivariate bicycle} codes is reported as a $\nkd{360, 12, \leq 24}$ code~\citep{bravyi2024high}, because constraint solvers time out before establishing a matching lower bound~\citep{webster2026distance}.
The problem compounds for code families, whose parameters are claimed for \emph{infinitely many} codes at once and cannot be checked by enumeration.
Meanwhile, new codes of increasing mathematical complexity are being proposed at an accelerating rate, increasingly by \textsc{ai}-driven search~\citep{olle2024simultaneous,su2025discovery,cruzbenito2026evolutionary,liu2026large,yan2026omniqec}.
Each new code typically arrives with its own formalism, notation, and proof techniques, and its claimed parameters often rest on informal arguments or simulation rather than complete proofs.
In this paper, we argue that this calls for a formal foundation in which codes are specified precisely and their parameters are proven, with machine-checked proofs that scale to large codes and to entire code families.
\begin{figure}[t]
\centering
\includegraphics[width=\linewidth]{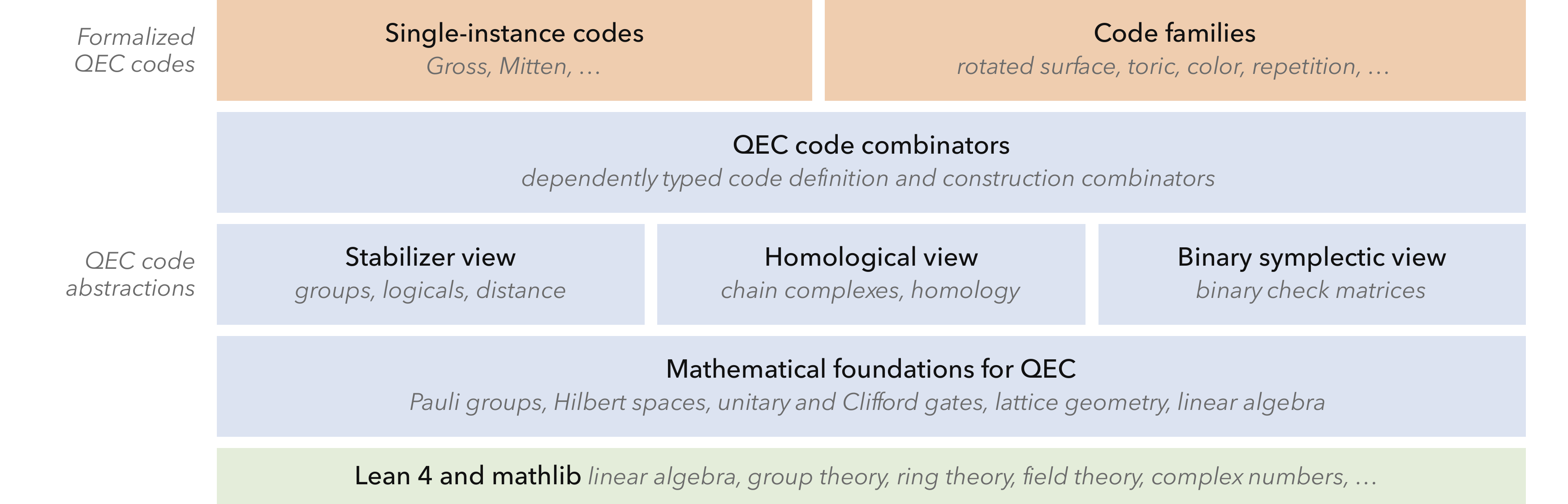}
\caption{Overview of \f's architecture}
\label{fig:architecture}
\end{figure}

\subsection{A Formal Foundation for Specifying and Verifying \textsc{qec} Codes}

This paper proposes \f, a Lean~\citep{moura2021lean4} framework for formally verified \textsc{qec} codes.
Formalizing \textsc{qec} codes has three key benefits. First, it establishes \emph{correctness}: every claimed code parameter is backed by a machine-checked proof. Second, it \emph{unifies} the diverse presentations of codes in the literature within a single framework in which codes are easy to specify. Third, it enables \emph{analytical} proofs of distance that are inexpressible using existing approaches.

Our verification framework, \cref{fig:architecture}, consists of (1) algebraic abstractions for specifying \textsc{qec} codes and code families in Lean as well as (2) proof techniques and algorithmic infrastructure for establishing encoding rate, distance, and the correctness of code construction. We evaluate our verification framework by formally verifying the properties of several prominent and recent \textsc{qec} codes and code families that are influencing the design and construction of quantum computers.

\paragraph{Algebraic Abstractions} \textsc{qec} codes are studied through the lens of several domains of mathematics. Each domain elucidates a different aspect of the code; the encoding rate may be easily understood and computed in one view, while distance falls out more easily in another. \f is organized as a body of abstractions to support these differing views of \textsc{qec} codes. In particular, we support the \emph{stabilizer view}, which we take to be the canonical specification for stabilizer codes, the \emph{binary symplectic view}, and the \emph{homological view}.
Lean's dependent type system lets us state and manipulate a code's parameters in its type: a value of type \lstinline|Code[[n, k, d]]| is a $\nkd{n,k,d}$  code together with machine-checked proofs that it uses $n$ physical qubits, encodes $k$ logical qubits, and has distance $d$.

The stabilizer view specifies a stabilizer code by a group of operators called stabilizer generators. Checking the correctness of the construction and the error-correcting properties of a stabilizer code amounts to proving algebraic properties of these stabilizer generators. Some of these proofs, when carried out from the definitions within the stabilizer view, lead to computationally intractable procedures for large codes. For example, proving that the stabilizer generators form an \emph{independent set} directly from the definition of independence requires reasoning about $2^{m - 1}$ products of generators for a set of $m$ generators. In the symplectic view of a stabilizer code, checking independence is a rank computation on an $m \times 2n$ matrix. This takes polynomial time by Gaussian elimination in contrast to the exponential approach derived from the definition in the stabilizer view.

The linear algebraic view, however, doesn't expose the right structural properties of the stabilizer generators to reason about code distance, a critical property of a code. For many important codes, homological algebra enables us to see structural properties of a code that the stabilizer and symplectic views miss out on. In some cases, homological algebra gives us enough theory to structurally deduce the distance of a code. For example, we formalize a proof that the toric code family defined over an $L \times L$ grid has distance $L$ parametrically via a homological argument.

In summary, because the views are connected by proven translations, researchers can specify a code once in the stabilizer view and prove each property in whichever view makes it tractable.
Specifying a code is lightweight: the user lists its stabilizer generators, and decision procedures discharge routine obligations (e.g., commutation), leaving only distance to genuine proof effort.

\paragraph{\textsc{qec} Code Combinators} In our work, we take a compositional view of distance computation, and we think of it through the lens of what we call \emph{\textsc{qec} code combinators}.
A combinator is a function which builds a new \textsc{qec} code from one or more existing ones, together with \emph{proof rules} that derive the parameters of the new code.
Dependent types make combinators natural to express in \f.

The simplest combinator is \emph{code concatenation}~\citep{knill1996concatenated,gottesman1997stabilizer}, a well-known approach that composes two codes by encoding each qubit of one code with the other. The distance of the concatenated code is at least the product of the two distances.
From a dependent types (in Lean) perspective, the type of this combinator can be cleanly stated as follows:
\begin{leancode}
def concat : Code[[n, k, d]] → Code[[n', k', d']] → Code[[n * n', k * k', ≥ d * d']]
\end{leancode}

While concatenation is a mechanical procedure for constructing new codes, most codes of practical interest are not constructed via concatenation.
We therefore propose a new combinator that generalizes an existing construction by \citet{symons2025sequences} to arbitrary stabilizer codes.
We call this the \emph{double cover} combinator, which consists of our generalized construction along with proof rules that determine the distance of the constructed code.

The construction glues two copies of a single base code into a new code called the \emph{double cover} with twice as many qubits. We propose two proof rules for this construction. The first provides a lower bound on the distance of the double cover. The second shows that, under conditions that can be checked on the base code, the distance exactly doubles. Using the first rule, we reduce the distance of the $\nkd{360, 12, \leq 24}$ code to a machine-checked search on a much smaller code and prove that it is exactly 24, previously known only as an upper bound.

\paragraph{Case Studies}  We formalize a series of \textsc{qec} codes and code families that are either canonical to the \textsc{qec} literature or are on the roadmap for being used as part of a leading fault-tolerant quantum computing architecture over the next few years. \Cref{tab:case-studies} lists some of the codes and code families we formalize and their parameters.
For example, we formalize the \emph{toric} code family and the \emph{triangular color} code family, both of which have recently been experimentally demonstrated on neutral-atom and ion-trap quantum computers~\citep{bluvstein2022quantum,bluvstein2024logical,salesrodriguez2025experimental,paetznick2024demonstration}. To our knowledge, these are the first machine-checked distance proofs for quantum code families, which no existing approach can establish. Additionally, we specify and verify individual codes, such as \textsc{ibm}'s \emph{gross} code, which is on \textsc{ibm}'s quantum computing roadmap~\citep{bravyi2024high,yoder2025tour}, and the recent \emph{mitten} code from  Oratomic~\citep{bhardwaj2026mitten}.

\subsection{Contributions} To summarize, this paper makes the following contributions:
\begin{itemize}
\item Dependently typed abstractions of three complementary algebraic formulations of \textsc{qec} codes and code families: stabilizer groups, symplectic representations, and chain complexes (\cref{sec:abstractions,sec:homological}).
\item Proof techniques for establishing the properties of \textsc{qec} codes and code families. Notably, we introduce \emph{code combinators} for compositionally establishing distance of \textsc{qec} codes (\cref{sec:distance}).
\item A Lean library including the theoretical foundations of \textsc{qec} codes, our abstractions and proof techniques, built heavily on Mathlib's group theory, linear algebra, and homology~\citep{mathlib2020}.
\item Case studies specifying and verifying codes and code families on the roadmaps of major quantum computing efforts. These include the first proof that \textsc{ibm}'s $\nkd{360, 12, 24}$ bivariate bicycle code has distance 24, previously known only as an upper bound, and the first mechanically checked distance proofs for infinite families of quantum codes, such as the toric, rotated surface, and triangular color code families (\cref{sec:case-studies}).
\end{itemize}

Our verification is restricted to algebraic constructions of codes and does not verify \emph{implementations} on a quantum computer. For instance, we do not verify the \emph{circuit-level} distance of a particular implementation of a code---a measure of how many errors the implementation can withstand. Instead, we verify the code distance, which is an algebraic property of the code. We argue that formally verifying the algebraic properties of codes is a starting point for a more comprehensive verification of a fault-tolerant system since it provides information that many parts of the quantum computing stack, such as resource estimation and logical compilation, rely on. Our work lays the foundations required to (1) reason about circuit-level implementations of these codes on given architectures and (2) build end-to-end verified compilers for fault-tolerant quantum computers.

\section{Quantum Error Correction: A Primer}\label{sec:qec}

In this section, we introduce the basics of \emph{quantum error correction} (\textsc{qec}) codes with a focus on \emph{stabilizer codes}~\citep{gottesman1997stabilizer,nielsen2010quantum}.
We will assume that the reader is familiar with basic concepts from quantum information theory: quantum states, quantum gates (unitaries), and measurement.

\subsection{Quantum Errors}
Error correction with a \textsc{qec} code proceeds in three steps: \emph{encoding} logical qubits redundantly into physical qubits, \emph{detecting} errors, and \emph{decoding}, which infers and undoes them. On classical computers, errors are bit flips: $0 \mapsto 1$, or $1 \mapsto 0$.
On quantum computers, the fundamental piece of information is the quantum bit (qubit), a superposition of the \emph{basis states}, $\ket{0}$ and $\ket{1}$, typically written as
$\alpha \ket{0} + \beta \ket{1},$ where $\alpha, \beta$ are complex \emph{amplitudes}.

Thus, errors on a quantum computer form a continuous space: an error can change $\alpha$ and $\beta$ by arbitrary amounts. Perhaps one of the most counterintuitive facts about quantum error correction is that we can \emph{discretize} the space of errors that we need to reason about~\citep{shor1995scheme,steane1996multiple,knill1997theory}. Specifically, to perform error correction, we only need to care about two kinds of \emph{Pauli errors}, $X$ and $Z$.
An $X$ error flips the basis states, and a $Z$ error negates the amplitude of $\ket{1}$:
\[
  X(\alpha \ket{0} + \beta \ket{1}) = \alpha \ket{1} + \beta \ket{0}, \qquad
  Z(\alpha \ket{0} + \beta \ket{1}) = \alpha \ket{0} - \beta \ket{1}.
\]
All unitary errors on a qubit can be captured as weighted sums and products of $X$ and $Z$ errors!

\begin{table}[t!]
\caption{Codes and code families formalized in \f and their $\nkd{n,k,d}$ parameters.}
\label{tab:case-studies}
\footnotesize
\begin{tabular}{@{}l>{\raggedright\arraybackslash}p{0.5\linewidth}@{}}
  \toprule
  Code (family) & Parameters \\
  \midrule
  Gross code (\textsc{bb} family) & $\nkd{144,\, 12,\, 12}$ \\
  \addlinespace[0.5em]
  $360$-qubit \textsc{bb} code & $\nkd{360,\, 12,\, 24}$; distance previously known only as $\leq 24$ \\
  \addlinespace[0.5em]
  Mitten code & $\nkd{150,\, 30,\, 10}$ \\
  \addlinespace[0.5em]
  Rotated surface code family & $\nkd{L^2,\, 1,\, L}$ on an $L \times L$ grid \\
  \addlinespace[0.5em]
  Toric code family & $\nkd{2L^2,\, 2,\, L}$ on an $L \times L$ periodic grid \\
  \addlinespace[0.5em]
  Triangular color code family & $\nkd{3r(r+1)+1,\, 1,\, 2r+1}$; $r=1$ is the $\nkd{7,1,3}$ Steane code \\
  \bottomrule
\end{tabular}
\end{table}

\subsection{Simple Example: The Bit-Flip Code}
We will first describe a simple quantum error-correcting code called the three-qubit bit-flip code~\citep{nielsen2010quantum} that is able to detect and correct a single $X$ error.

\paragraph{Encoding a Logical Qubit}

\begin{figure}[t]
\centering
\footnotesize
\begin{quantikz}[column sep=0.55cm, row sep={0.6cm,between origins}]
  \lstick{$\q{q_1}$}
    & \ctrl{1}\gategroup[3,steps=2,style={draw=none, fill=ccgreen, fill opacity=0.4, rounded corners}, background]{encode} & \ctrl{2} & \gate[style={draw=maroon, dashed}, label style={maroon}]{X}
    & \ctrl{3}\gategroup[5,steps=4,style={draw=none, fill=zstab, fill opacity=0.4, rounded corners}, background]{syndrome extraction} & \qw & \qw & \qw & \qw \\
  \lstick{$\q{q_2}$}
    & \targ{} & \qw & \qw
    & \qw & \ctrl{2} & \ctrl{3} & \qw & \qw \\
  \lstick{$\q{q_3}$}
    & \qw & \targ{} & \qw
    & \qw & \qw & \qw & \ctrl{2} & \qw \\
  \lstick{$\q{a_1} = \ket{0}$}
    & \qw & \qw & \qw
    & \targ{} & \targ{} & \qw & \qw & \meter{} \\
  \lstick{$\q{a_2} = \ket{0}$}
    & \qw & \qw & \qw
    & \qw & \qw & \targ{} & \targ{} & \meter{}
\end{quantikz}
\caption{Encoding, a single bit-flip error, and syndrome extraction for the three-qubit bit-flip code. The \emph{encode} block prepares $\alpha\ket{000} + \beta\ket{111}$ with two \textsc{cnot} gates; the dashed $X$ marks an error on qubit 1; the \emph{syndrome extraction} block copies $q_1 \oplus q_2$ and $q_2 \oplus q_3$ into the ancillas, which are then measured to obtain the syndrome.}
\label{fig:bitflip-circuit}
\end{figure}
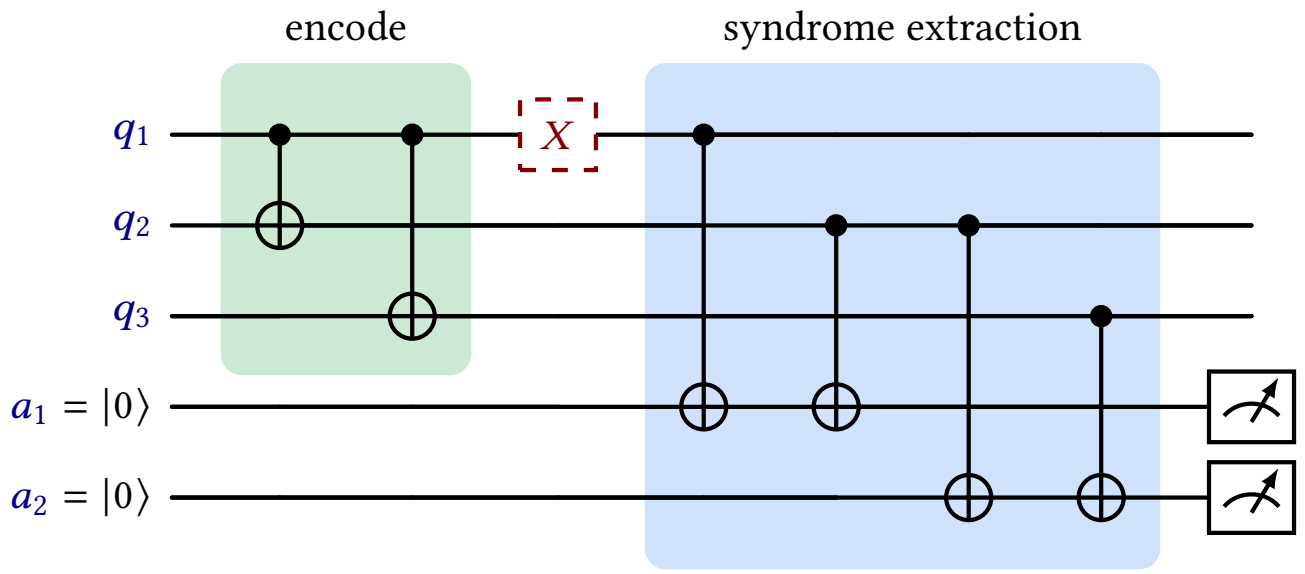

The key idea is that we encode a single qubit state $\alpha \ket{0} + \beta \ket{1}$ using three qubits as $\alpha \ket{000} + \beta \ket{111}$. This encoding can be done with two \textsc{cnot} gates, shown in the \emph{encode} block of the circuit in \Cref{fig:bitflip-circuit}.

We say that the three \emph{physical qubits} encode a single \emph{logical qubit}, which serves as an abstraction of a qubit that we're later able to program with. In particular, the logical $\ket{0}$ state, which we will denote $\ket{0_L}$, will be encoded by the physical quantum state $\ket{000}$. Similarly, the logical $\ket{1_L}$ state will be encoded by the physical quantum state $\ket{111}$.

\[
  \ket{0_L} \equiv \ket{000} \qquad \ket{1_L} \equiv \ket{111} \qquad \ket{\pm_L} \equiv \tfrac{1}{\sqrt{2}}\left(\ket{000} \pm \ket{111}\right)
\]

\begin{wraptable}[12]{r}{0.25\linewidth}
\centering
\caption{Syndromes of the three-qubit bit-flip code and the correction applied for each.}
\label{tab:rep-syndromes}
\begin{tabular}{ccl}
  \toprule
  $a_1$ & $a_2$ & Correction \\
  \midrule
  0 & 0 & none \\
  1 & 0 & $X_1$ \\
  1 & 1 & $X_2$ \\
  0 & 1 & $X_3$ \\
  \bottomrule
\end{tabular}
\end{wraptable}
\paragraph{Error Detection}
Suppose that a single $X$ error occurs on the first of the three physical qubits.
This changes the state from $\alpha \ket{000} + \beta \ket{111}$ to
$\alpha \ket{100} + \beta \ket{011}.$
We can easily detect this error by observing that the basis states are no longer $\ket{000}$ and $\ket{111}$.
However, we cannot simply measure the qubits to check this, since measuring would collapse the superposition and destroy $\alpha$ and $\beta$.
It turns out that we can instead detect such an error by computing parities of pairs of qubits, which reveal where the error is without revealing $\alpha$ and $\beta$.
Specifically, the parities $q_1 \oplus q_2$ and $q_2 \oplus q_3$ are both $0$ on $\ket{000}$ and $\ket{111}$, but after the error, both $\ket{100}$ and $\ket{011}$ have $q_1 \oplus q_2 = 1$ and $q_2 \oplus q_3 = 0$.
This pair of parities identifies the flipped qubit, as \cref{tab:rep-syndromes} shows for each single $X$ error.

We compute the two parities into ancillary qubits $a_1$ and $a_2$ with \textsc{cnot} gates and measure them---see the \emph{syndrome extraction} block of \Cref{fig:bitflip-circuit}. The outcomes, called the \emph{syndrome}, tell us whether a single $X$ error occurred and on which qubit; we correct it by applying $X$ to that qubit (\cref{tab:rep-syndromes}).

This error-correcting code can only detect and correct $X$ errors, but fails to detect (or correct) $Z$ errors. Moreover, it can only correct $X$ errors if at most one qubit is affected at a time. \textsc{qec} studies schemes like the three-qubit bit-flip code to correct both kinds of errors as well as handle multiple qubit errors while trying to minimize the amount of redundancy in the encoding. Describing these codes concisely requires a more abstract view of the syndrome extraction step.

\subsection{The Stabilizer View}
\paragraph{Stabilizer Generators}
Syndrome extraction is an instance of a more general primitive: \emph{measuring an operator}. In this language, the circuit in \Cref{fig:bitflip-circuit} measures the operators $Z_1Z_2$ and $Z_2Z_3$, where $Z_1Z_2 = Z \otimes Z \otimes I$ applies $Z$ to qubits 1 and 2. Measuring $Z_1Z_2$ yields the eigenvalue $(-1)^{q_1 \oplus q_2}$: $+1$ on computational basis states where qubits 1 and 2 agree and $-1$ where they differ, so it measures their parity; likewise, $Z_2Z_3$ measures the parity of qubits 2 and 3. The bit read from each ancilla is the exponent, so outcome $0$ corresponds to $+1$ and outcome $1$ to $-1$. We call $Z_1Z_2$ and $Z_2Z_3$ the \emph{stabilizer generators} of the code. Codes described by the stabilizer generators they measure are called \emph{stabilizer codes}~\citep{gottesman1996class,calderbank1997quantum}; they underpin most fault-tolerant architectures.

\paragraph{Codespace}
The stabilizer generators determine the \emph{codespace} of the code: the states on which every generator measures $+1$, that is, the intersection of their $+1$-eigenspaces. For the bit-flip code, these are the states in which qubits 1 and 2 agree and qubits 2 and 3 agree, so the codespace is spanned by $\ket{000}$ and $\ket{111}$, which are the encoded states $\alpha \ket{000} + \beta \ket{111}$. In general, a code on $n$ physical qubits encodes $k$ logical qubits when its codespace has dimension $2^k$; here, $n = 3$ and $k = 1$. An error that moves the state out of the codespace makes at least one generator measure $-1$. The list of outcomes is the syndrome displayed in \cref{tab:rep-syndromes}.

\paragraph{Logical Operators and Distance}
Not every error leaves the codespace, however. A $Z$ error on the first qubit maps $\alpha \ket{000} + \beta \ket{111}$ to $\alpha \ket{000} - \beta \ket{111}$, which is still in the codespace, so every generator still measures $+1$ and the error goes undetected. We call an operator that maps every state in the codespace to a state in the codespace, and so goes undetected, a \emph{logical operator}; the stabilizer generators are logical operators too, but trivial ones, since they act as the identity on the codespace. We call a logical operator like $Z_1$, which changes the encoded state without being detected, a \emph{nontrivial logical operator}, and the resulting error a \emph{logical error}.

A robust code has only nontrivial logical operators that are unlikely to arise as physical errors, which we measure by their \emph{weight}: the number of qubits on which an operator acts as something other than the identity. The \emph{distance} $d$ of a code is the minimum weight of a nontrivial logical operator (Definition~\ref{def:distance} makes this precise), that is, the fewest single-qubit errors that can cause an undetectable logical error. Together with $n$ and $k$, the distance gives the standard $\nkd{n, k, d}$ description of a code~\citep{calderbank1998quantum}. Since $Z_1$ has weight $1$, the bit-flip code has distance $1$ and is a $\nkd{3, 1, 1}$ code.

\subsection{Code Families}\label{sec:families}
In practice, many \textsc{qec} codes come as \emph{code families}: an infinite set of \textsc{qec} codes that share similar structure.
For example, we can generalize the three-qubit bit-flip code to an $n$-qubit one, resulting in the family of codes of the form $\nkd{n, 1, 1}$ for $n \geq 2$.
Notice that as we increase $n$, here the number of logical qubits and distance remain 1. In general, however, we have powerful code families in which one or both quantities increase with more physical qubits.

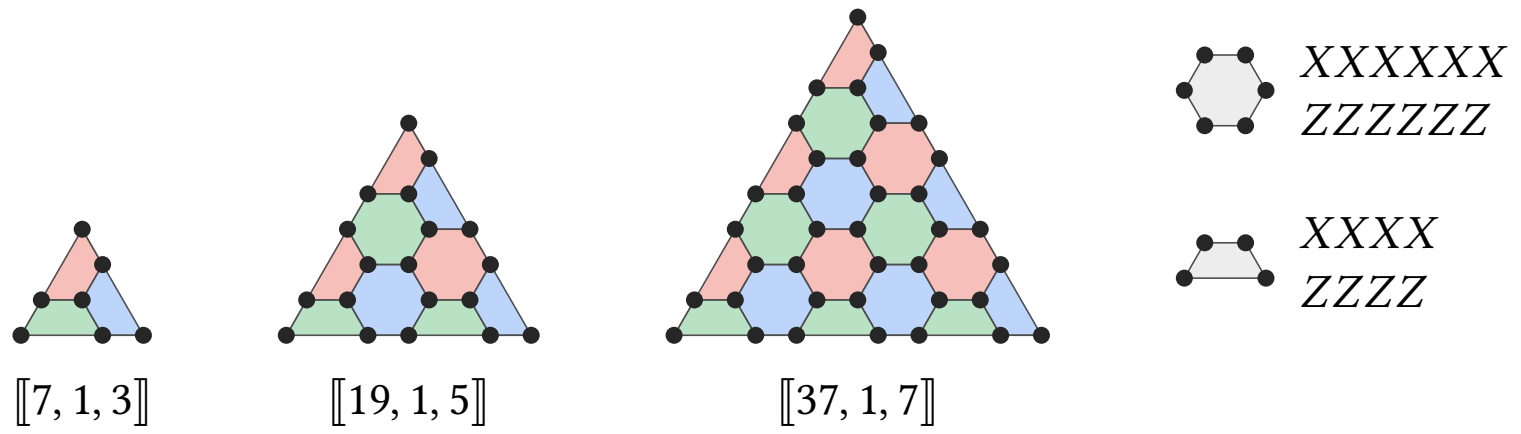
\begin{figure}[t]
\centering
\begin{tikzpicture}[x=0.27cm, y=0.27cm]
  \foreach \d/\x in {3/0, 5/6.5, 7/16} {
    \begin{scope}[shift={(\x,0)}]
      \colorcodepatch{\d}
      \pgfmathtruncatemacro{\n}{(3*\d*\d+1)/4}
      \node[font=\small, anchor=north] at ({3*(\d-1)/4},-0.6) {$\nkd{\n,1,\d}$};
    \end{scope}
  }
  \begin{scope}[shift={(29.5,0)},
                q/.style={fill=black!85}, lbl/.style={font=\small, anchor=west, align=left}]
    \filldraw[chk, fill=black!12] (0,6) +(0:1) -- +(60:1) -- +(120:1) -- +(180:1)
      -- +(240:1) -- +(300:1) -- cycle;
    \foreach \t in {0,60,...,300} \fill[q] (0,6) +(\t:1) circle[radius=1.6pt];
    \node[lbl] at (1.4,6) {$XXXXXX$\\$ZZZZZZ$};
    \filldraw[chk, fill=black!12] (0,1.4) +(0:1) -- +(60:1) -- +(120:1) -- +(180:1) -- cycle;
    \foreach \t in {0,60,120,180} \fill[q] (0,1.4) +(\t:1) circle[radius=1.6pt];
    \node[lbl] at (1.4,1.8) {$XXXX$\\$ZZZZ$};
  \end{scope}
\end{tikzpicture}
\caption{Triangular color code patches; the $d=3$ patch is the $\nkd{7,1,3}$ Steane code~\citep{steane1996error}. \textbf{Qubits are the vertices}; each face carries an $X$ and a $Z$ generator on all of its qubits.
}
\label{fig:color-patches}
\end{figure}

\paragraph{Color Code Family}
To give some intuition, consider the \emph{color code family}~\citep{bombin2006topological,landahl2011fault} shown in \cref{fig:color-patches}. This is the family of codes of the form:
\[
  \nkd{\underbrace{3r(r+1)+1}_{n},\, \underbrace{1}_{k},\, \underbrace{2r+1}_{d}} \qquad \text{for } r \geq 1.
\]
Observe how the distance increases with more physical qubits.
Here, the qubits are laid out geometrically as a triangle: each qubit is a vertex of a tiling of the triangle into faces, and the stabilizer generators act on the qubits around each face. For example, each hexagonal face represents two stabilizer generators, $XXXXXX$ and $ZZZZZZ$ on its six qubits, while each four-sided face along the boundary represents $XXXX$ and $ZZZZ$ on its four qubits.

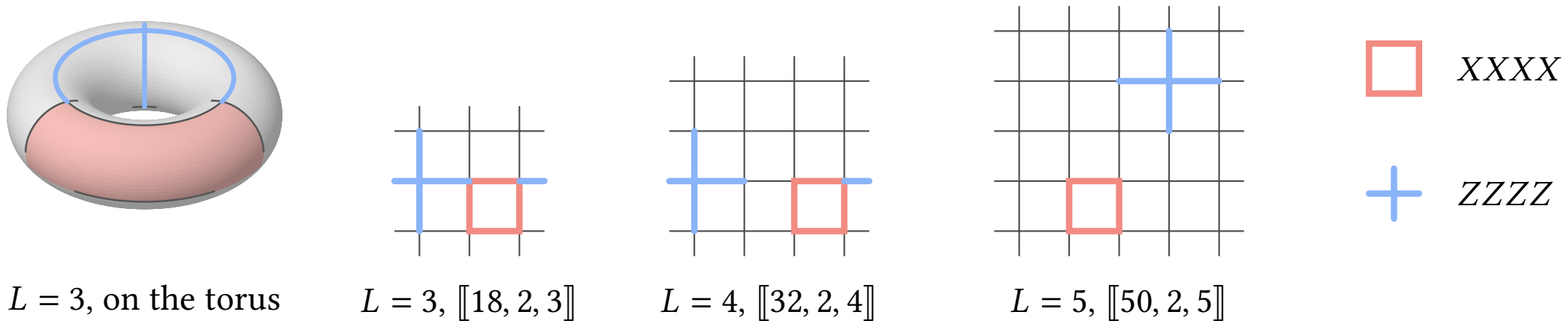
\begin{figure}[t]
\centering
\begin{tikzpicture}[x=0.44cm, y=0.44cm]
  \node[anchor=south, inner sep=0] at (2.5,0)
    {\includegraphics[width=2.5cm]{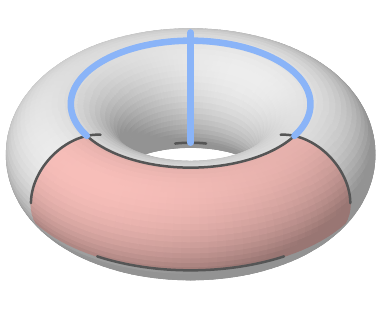}};
  \node[font=\small, anchor=north] at (2.5,-0.8) {$L=3$, on the torus};
  \foreach \L/\fc/\vx/\vy/\x in {3/1/0/1/8, 4/2/0/1/13.5, 5/1/3/3/20} {
    \begin{scope}[shift={(\x,0)}]
      \toricgrid{\L}{\fc}{\vx}{\vy}
      \pgfmathtruncatemacro{\n}{2*\L*\L}
      \node[font=\small, anchor=north] at ({(\L-1)/2},-0.8) {$L=\L$, $\nkd{\n,2,\L}$};
    \end{scope}
  }
  \begin{scope}[shift={(27,0)},
                lbl/.style={font=\small, anchor=west}]
    \draw[ccred, line width=1.6pt, line cap=round] (0,2.75) rectangle ++(1,1);
    \node[lbl] at (1.5,3.25) {$XXXX$};
    \draw[ccblue, line width=1.6pt, line cap=round] (0,0.75) -- (1,0.75) (0.5,0.25) -- (0.5,1.25);
    \node[lbl] at (1.5,0.75) {$ZZZZ$};
  \end{scope}
\end{tikzpicture}
\caption{Toric code lattices: $L=3$ on the torus (left) and $L=3,4,5$ cut open into grids. \textbf{Qubits are the edges}; stubs are halves of edges that wrap to the opposite side. Each grid shows one face generator ($X$, red) and one vertex generator ($Z$, blue), which wraps around for $L=3,4$.}
\label{fig:toric-grids}
\end{figure}

\paragraph{Toric Code Family}
Now let us consider a different code family, the \emph{toric code}~\citep{kitaev2003fault}. Here the geometry is different: qubits are laid out as a grid on the surface of a torus (a donut).
\cref{fig:toric-grids} (left) shows one such torus.
As we grow the torus, both in diameter and thickness, by adding physical qubits, we grow code distance, giving codes of the form $\nkd{2L^2, 2, L}$ for $L \geq 2$.

Observe that here we encode two logical qubits.
For visualization, we cut the torus open and flatten it into an $L \times L$ grid, as shown in \cref{fig:toric-grids} (right).
Because the grid comes from a torus, its opposite sides are identified: an edge leaving the right (top) side continues on the left (bottom) side.
The qubits are the edges of the grid---one horizontal and one vertical edge per vertex, $2L^2$ in total.
There are two kinds of stabilizer generators: \emph{face} generators apply $X$ to the four edges around a face (red), and \emph{vertex} generators apply $Z$ to the four edges meeting at a vertex (blue).

\section{Specifying \textsc{qec} Codes}\label{sec:abstractions}

In this section, we describe \f's view of \textsc{qec} codes, how our abstractions support the construction of codes, and techniques we use to establish code parameters. We begin by describing the stabilizer formalism, which gives us the language to describe an important class of codes called stabilizer codes---the main object of study for \f.

\subsection{Stabilizer Formalism}\label{sec:stabilizer}

In \cref{sec:qec}, we described the syndrome extraction step for the bit-flip code and then used the language of operator measurement to describe it as measuring the stabilizer generators $Z_1Z_2$ and $Z_2Z_3$. The point is that we could describe the code completely by the stabilizer generators that are measured. The definition of a \emph{stabilizer code} tells us precisely which algebraic properties the stabilizers need to satisfy for the induced code to be able to detect and correct errors. These properties are stated in the language of group theory, so we start there.

\subsubsection{Pauli Groups}
First, let us recall the Pauli operators, $I, X, Y, Z$:
\[
  I = \begin{pmatrix} 1 & 0 \\ 0 & 1 \end{pmatrix} \qquad
  X = \begin{pmatrix} 0 & 1 \\ 1 & 0 \end{pmatrix} \qquad
  Y = \begin{pmatrix} 0 & -i \\ i & 0 \end{pmatrix} \qquad
  Z = \begin{pmatrix} 1 & 0 \\ 0 & -1 \end{pmatrix}
\]

The Pauli group $P_n$ on $n$ qubits is the set of $n$-fold tensor products of Pauli matrices with a phase:

\begin{definition}[Pauli group]\label{def:pauli}
The \emph{Pauli group} on $n$ qubits is the group, under matrix multiplication,
\[
  P_n = \{i^k\, Q_1 \otimes \cdots \otimes Q_n : Q_j \in \{I, X, Y, Z\}\}.
\]
\end{definition}

\noindent We call the elements of $P_n$ \emph{Pauli operators}, or simply \emph{operators}. The \emph{weight} $\mathrm{wt}(P)$ of $P \in P_n$ is the number of qubits on which it acts as a Pauli other than $I$; for example, $\mathrm{wt}(Z \otimes Z \otimes I) = 2$. Any two elements $P, Q \in P_n$ either \emph{commute} ($PQ = QP$) or \emph{anticommute} ($PQ = -QP$).

Our corresponding Lean formalization of elements of $P_n$ is below:
\begin{leancode}
structure NPauliOperator (n : ℕ) where
  phasePower : Fin 4
  operators : Fin n → PauliOperator -- I, X, Y, Z on each qubit
\end{leancode}
\lstinline{phasePower} defines the power of $i$ (we only need $k \in [0,3]$ because $i^4 = 1$, hence the type \lstinline{Fin 4})
and \lstinline{operators} defines each of the $Q_i$ operators in \cref{def:pauli}.
We define the multiplication operation for these elements and prove that \lstinline{NPauliOperator n} forms a group using Mathlib's definition of a group. Writing operators as structure instances in Lean becomes unwieldy, so \f provides the notation \lstinline{σ[ZZI]} for $Z_1Z_2 := Z \otimes Z \otimes I$. The two stabilizer generators of the bit-flip code we described are then \lstinline{σ[ZZI]} and \lstinline{σ[IZZ]}.

\subsubsection{Stabilizer Codes}
The theory of stabilizer codes~\citep{gottesman1996class,calderbank1997quantum,gottesman1997stabilizer} tells us the precise conditions that a set of operators in $P_n$ needs to satisfy in order to define a valid code.

\begin{definition}[stabilizer group]\label{def:stabilizer-group}
A \emph{stabilizer group} is a subgroup $G \subset P_n$ such that (1) any two of its elements commute (\emph{abelian}) and (2) $-I \notin G$.
\end{definition}

\begin{definition}[stabilizer code]\label{def:stabilizer-code}
A stabilizer group $G$ is an $\nkd{n, k}$ \emph{stabilizer code} if $k \le n$ and $|G| = 2^{n-k}$.
Its \emph{codespace} is the common $+1$-eigenspace, i.e.,
$\{\ket{\psi} : g \ket{\psi} = \ket{\psi} \text{ for all } g \in G\}$.
\end{definition}

In \f, we formalize these definitions of stabilizer groups and stabilizer codes as below. \lstinline{StabilizerCode n k} is the central abstraction in the library since it specifies what it means to be a valid stabilizer code. We write \lstinline{Code[[n, k]]} as custom notation for \lstinline{StabilizerCode n k}.
\begin{leancode}
structure StabilizerGroup (n : ℕ) where
  toSubgroup : Subgroup (NPauliOperator n)
  is_abelian : ∀ g h, g ∈ toSubgroup → h ∈ toSubgroup → g * h = h * g
  no_neg_identity : (*$-I_n$*) ∉ toSubgroup

structure StabilizerCode (n k : ℕ) extends StabilizerGroup n where
  hk : k ≤ n
  card_eq : Nat.card toSubgroup = 2 ^ (n - k)
\end{leancode}
\lstinline{Subgroup} is Mathlib's notion of a subgroup, and $-I_n$ is $-I$ on $n$ qubits. Since every element of a stabilizer group squares to $I$, one can show that having $n-k$ independent stabilizer generators in a stabilizer group $G$ is equivalent to $|G| = 2^{n-k}$. \f therefore defines a stabilizer code to be a stabilizer group $G$ with $|G| = 2^{n-k}$. \f  also proves that the codespace of a \lstinline{StabilizerCode n k} is a complex vector space of dimension $2^k$ (i.e., it encodes $k$ logical qubits).

\subsubsection{Constructing Stabilizer Codes}
The size of the stabilizer group $G$ defining a stabilizer code is exponential in $n$ and so we cannot state the elements of $G$ explicitly in general.
Luckily, we can use a small set of elements, \emph{generators}, to specify the entire group.
A set $S \subseteq P_n$ \emph{generates} the group $\langle S \rangle$ of all products of elements of $S$; we call the elements of $S$ \emph{generators}.
$S$ is \emph{independent} if no generator is a product of the others.
For the bit-flip code, the two generators $Z_1Z_2, Z_2Z_3$ result in the stabilizer group $\{I, Z_1Z_2, Z_2Z_3, Z_1Z_3\}$.

In \f, \textsc{qec} codes can be constructed from a \emph{presentation}: a list of generators together with the proofs of structural properties required of a stabilizer code. We then show that an instance of a \lstinline{StabilizerCode.Presentation n k} indeed does meet the specifications of a \lstinline{StabilizerCode}. This is detailed in the \lstinline{StabilizerCode.Presentation.toCode} definition, for which we show only the type signature below, for the sake of brevity.

\begin{leancode}
structure StabilizerCode.Presentation (n k : ℕ) where
  hk : k ≤ n
  generatorsList : List (NPauliOperator n)
  generators_length : generatorsList.length = n - k
  generators_phaseZero : AllPhaseZero generatorsList
  generators_independent : rowsLinearIndependent generatorsList
  generators_commute : ∀ g,h ∈ generatorsList, g * h = h * g
  closure_no_neg_identity : (*$-I_n$*) ∉ Subgroup.closure (generatorsList) := ...

noncomputable def StabilizerCode.Presentation.toCode
    (P : StabilizerCode.Presentation n k) : StabilizerCode n k
\end{leancode}

 The first three fields of \lstinline{StabilizerCode.Presentation} define a list of generators.
The fourth field requires that all generators have phase power zero (i.e.\ phase $+1$) without loss of generality so that we can use the binary symplectic representation (\cref{sec:symplectic}). We define independence of the list of generators as linear independence of the generators' rows in the binary symplectic representation. The last field states that the closure of the generators (using Mathlib's \lstinline{Subgroup.closure}) does not contain $-I$. We also provide a default proof of this (omitted above) since it follows from all the previous fields.

For the bit-flip code, every obligation is discharged by the \lstinline{decide} tactic, which runs a decision procedure through the Lean kernel.

\begin{leancode}
def bitFlipPresentation : StabilizerCode.Presentation 3 1 where
  hk := by decide
  generatorsList := [σ[ZZI], σ[IZZ]]
  generators_length := by decide
  ...

noncomputable def bitFlipCode : Code[[3, 1]] := bitFlipPresentation.toCode
\end{leancode}

\subsubsection{Code Families}
The power of Lean especially shines when we consider code families, where we can define an infinite set of stabilizer codes parametrically. For instance, we can extend the three-qubit code from earlier to an $(n+2)$-qubit code, for any $n$, by measuring the stabilizer generators $Z_iZ_{i+1}$ for $i = 0, \ldots, n$:

\begin{leancode}
def ZPair (n : ℕ) (i : Fin (n + 1)) : NPauliOperator (n + 2) :=
  σ[n + 2 | i ↦ Z, i + 1 ↦ Z]

def RepetitionCodeN.presentation (n : ℕ) : StabilizerCode.Presentation (n + 2) 1 where
  hk := by omega
  generatorsList := List.ofFn (ZPair n)
  ...

noncomputable def RepetitionCodeN.stabilizerCode (n : ℕ) : Code[[n + 2, 1]] :=
  (RepetitionCodeN.presentation n).toCode
\end{leancode}

\noindent \lstinline{ZPair n i} defines the $i$th stabilizer generator $Z_iZ_{i+1}$ for the $(n+2)$-qubit code. \f's notation \lstinline{σ[n + 2 | i ↦ Z, j ↦ Z]} is shorthand for the $n+2$ qubit operator that acts as $Z$ on qubits $i$ and $j$ and as $I$ elsewhere, and so $Z_iZ_{i+1}$ is expressed as \lstinline{σ[n + 2 | i ↦ Z, i + 1 ↦ Z]}. Because the generator list now depends on $n$, \lstinline{decide} no longer applies, and so each field is proven by a structural lemma instead: the generators commute because they are all $Z$-type, $-I \notin G$ follows from a general lemma for \textsc{css} codes~\citep{calderbank1996good,steane1996multiple} (i.e., stabilizer codes whose generators are each purely $X$- or $Z$-type), and independence is proven by induction over the columns of the check matrix.

\subsubsection{Example: The Toric Code}
We can specify the toric code family~\citep{kitaev2003fault} described in \cref{sec:families} (\cref{fig:toric-grids}) by the same mechanism. For an $L \times L$ periodic grid, \lstinline{hEdge x y} is the horizontal edge leaving the vertex $(x, y)$ and \lstinline{vEdge x y} is the vertical edge leaving it. The vertex and face generators are written with the same \lstinline{σ[n | i ↦ P, …]} notation as before:

\begin{leancode}
variable {L : ℕ} [Fact (0 < L)]   -- n := 2 * L * L

def vertexStab (x y : Fin L) : NPauliOperator n :=
  σ[n | hEdge x y ↦ Z, hEdge (x - 1) y ↦ Z,
        vEdge x y ↦ Z, vEdge x (y - 1) ↦ Z]

def faceStab (x y : Fin L) : NPauliOperator n :=
  σ[n | hEdge x y ↦ X, hEdge x (y + 1) ↦ X,
        vEdge x y ↦ X, vEdge (x + 1) y ↦ X]
\end{leancode}

\noindent \lstinline{vertexStab x y} applies $Z$ to the four edges incident to the vertex $(x, y)$, and \lstinline{faceStab x y} applies $X$ to the four edges bounding the face whose lower-left corner is $(x, y)$.

With this construction, the stabilizer generators of the toric code are \lstinline{vertexStab x y} and \lstinline{faceStab x y} with $(x, y)$ ranging over the grid.\footnote{Since every edge has two endpoints and borders two faces, the vertex generators multiply to $I$, as do the face generators. We therefore drop the two generators at the origin, leaving $2L^2 - 2$ independent generators and $k = 2$.}

\begin{leancode}
def toricPresentation (L : ℕ) [Fact (2 ≤ L)] :
    StabilizerCode.Presentation (2 * L * L) 2 where
  -- vertexStab x y and faceStab x y for every (x, y) ≠ (0, 0)
  generatorsList := generatorsListPackaged L
  ... -- proof obligations elided

noncomputable def toricStabilizerCode (L : ℕ) [Fact (2 ≤ L)] :
    Code[[2 * L * L, 2]] := (toricPresentation L).toCode
\end{leancode}

\noindent Like the bit-flip family, each obligation is discharged by a structural lemma. The homological view (\cref{sec:homological}) discusses how some of these proofs work.

\subsection{Binary Symplectic View}\label{sec:symplectic}
In the previous section, we discussed how \f's \lstinline{StabilizerCode.Presentation n k} abstraction allows us to construct concrete codes by specifying a list of stabilizer generators and proving key properties about them. The binary symplectic view gives us a straightforward and efficient way to discharge the independence and $-I$ obligations via linear algebra.

\paragraph{Operators as Binary Vectors} Ignoring phase, each qubit of a Pauli operator carries one of $I$, $X$, $Z$, or $Y$, and since $Y = XZ$ up to a phase, each is a product $X^xZ^z$ with two bits $(x, z)$: $x$ says whether an $X$ is present and $z$ whether a $Z$ is present.
\[
I \mapsto (0,0) \qquad X \mapsto (1,0) \qquad Z \mapsto (0,1) \qquad Y \mapsto (1,1)
\]
Collecting the $x$ bits of all $n$ qubits, then the $z$ bits, turns an operator $p \in P_n$ into a binary vector $(x_1 \ldots x_n \mid z_1 \ldots z_n) \in \mathbb{F}_2^{2n}$, called its \emph{symplectic vector}~\citep{calderbank1997quantum,calderbank1998quantum}.
For example, $p = ZXY$ has $X$ parts on qubits 2 and 3 and $Z$ parts on qubits 1 and 3, so its symplectic vector is $(011 \mid 101)$.

The \emph{quantum parity-check matrix} (referred to as \emph{check matrix} henceforth)~\citep{gottesman1997stabilizer} is the $(n-k) \times 2n$ matrix whose rows are the symplectic vectors of the $n-k$ stabilizer generators. The left $n$ columns correspond to the $X$ component of the operators and the right $n$ columns correspond to the $Z$ component. For example, the $\nkd{4, 2, 2}$ code has stabilizer generators $XXXX$ and $ZZZZ$ (left), with the $2 \times 8$ check matrix on the right:
\[
\setlength{\arraycolsep}{2.5pt}
\begin{array}{l}
  XXXX\\ ZZZZ
\end{array}
\quad\longmapsto\quad
\left(\begin{array}{cccc|cccc}
  1&1&1&1 & 0&0&0&0\\
  0&0&0&0 & 1&1&1&1
\end{array}\right)
\]
Independence of the stabilizer generators is the linear independence of the corresponding set of symplectic vectors, which is established by checking whether the corresponding check matrix has full row rank. For concrete codes, like the $\nkd{4, 2, 2}$ code, this can be done via a kernel-checkable procedure in polynomial time. For code families, this can be done via a structural argument.

Unfortunately, the symplectic view does not give us a way to efficiently compute the distance of the code, which is in general an \textsc{np}-hard problem~\citep{kapshikar2023hardness}. The homological view, discussed next, allows us to exploit the algebraic structure of a class of \textsc{qec} codes at a finer level to enable distance proofs that are parametric for a family or scale to large codes.
\section{Code Distance and the Homological View}

We now discuss our formalization of code distance, describe the homological view of \textsc{qec} codes, and discuss how it can give us the structural insights to prove distance for an entire code family.

\subsection{Code Distance}\label{sec:distance-def}

In \cref{sec:qec}, we described logical operators in terms of the codespace: an error goes undetected when it maps every codespace state to a codespace state, and it is a logical error when it moreover changes the encoded state. The stabilizer formalism gives us an algebraic condition for logical operators~\citep{gottesman1997stabilizer}. Recall that any two operators either commute or anticommute. If an operator $h$ commutes with a generator $g$, then $g(h\ket{\psi}) = h(g\ket{\psi}) = h\ket{\psi}$ for every codespace state $\ket{\psi}$, so $g$ still measures $+1$ after $h$ is applied. If $h$ anticommutes with $g$, the same measurement gives $-1$, and so $g$ \emph{detects} $h$. An error therefore goes undetected  when it commutes with every stabilizer. The set of such operators that commute with the stabilizers is called the \emph{centralizer} of $G$. Among them, the stabilizers themselves act as the identity on the codespace. The remaining elements of the centralizer are the ones that change the encoded state.

\begin{definition}[logical operators]\label{def:logical}
Let $G$ be a stabilizer code on $n$ qubits. The \emph{centralizer} of $G$ is $C(G) = \{h \in P_n : hg = gh \text{ for all } g \in G\}$, the set of operators that commute with every stabilizer, and its elements are called the \emph{logical operators} of $G$. A logical operator $A$ is \emph{nontrivial} if $i^k A \notin G$ for every $k \in \mathbb{N}$.
\end{definition}

An error operator $E$ is undetected by the stabilizers when it is a logical operator. However, in the case where $E$ is a stabilizer itself it doesn't change the encoded state, so $E$ is harmless. But if $E$ is a nontrivial logical operator, then it changes the encoded state while being undetected by the stabilizer generators. This is a logical error. So, a natural parameter to describe the error-detection capabilities of a \textsc{qec} code is the minimum weight of a logical error~\citep{calderbank1998quantum,gottesman1997stabilizer}.

\begin{definition}[distance]\label{def:distance}
The \emph{distance} of a stabilizer code $G$ is
\[
  d(G) = \min \{\mathrm{wt}(A) : A \text{ is a nontrivial logical operator of } G\},
\]
and an $\nkd{n, k, d}$ code is an $\nkd{n, k}$ stabilizer code of distance $d$.
\end{definition}

\noindent In \f, we define distance and an $\nkd{n, k, d}$ stabilizer code as follows:

\begin{leancode}
noncomputable def centralizer (S : StabilizerGroup n) : Subgroup (NPauliOperator n) :=
  Subgroup.centralizer (S.toSubgroup : Set (NPauliOperator n))

def IsNontrivial (g : NPauliOperator n) (S : StabilizerGroup n) : Prop :=
  g ∈ centralizer S ∧ ∀ s ∈ S.toSubgroup, s.operators ≠ g.operators

def HasCodeDistance (C : StabilizerCode n k) (d : ℕ) : Prop :=
  d ≥ 1 ∧ (∀ g, IsNontrivial g C.toStabilizerGroup → 0 < weight g → weight g ≥ d) ∧
  (∃ g, IsNontrivial g C.toStabilizerGroup ∧ weight g = d)

structure StabilizerCodeWithDistance (n k d : ℕ) extends StabilizerCode n k where
  hasDistance : HasCodeDistance toStabilizerCode d
\end{leancode}

\noindent For the definition of a nontrivial logical operator, we use Mathlib's \lstinline{Subgroup.centralizer}. A code bundled with a proof of its distance has type \lstinline{StabilizerCodeWithDistance n k d}, for which we write \lstinline{Code[[n, k, d]]}. For the bit-flip code, the logical operator \lstinline{σ[ZII]} of weight $1$ is a witness of distance $1$, and \f proves that the code is a \lstinline{Code[[3, 1, 1]]}.

\paragraph{Brute-Force Distance Computation}

What happens when we try to compute the code distance directly from this definition? The na\"ive approach is to follow \cref{def:distance} and enumerate all Paulis on the $n$ physical qubits in order of weight, until a nontrivial logical operator is found.

There are $\binom{n}{w} 3^w$ phase-free Paulis of weight $w$, so finding a distance of $d$ examines $\sum_{w \leq d} \binom{n}{w} 3^w \leq (3n)^d$ candidates, each needing a commutation test and a membership test that are polynomial in $n$ in the symplectic view.
Practical tools such as \textsc{ilp} and \textsc{sat} solvers prune this search but remain exponential in the worst case.
For example, in a recent benchmark of exact methods including \textsc{mip} and \textsc{sat} solvers, no method certified the distance of any quantum code above $d = 18$~\cite{webster2026distance}.

In addition to not being able to scale to codes with high distances, these techniques cannot establish the distance for a parametric family of codes, whose useful members often lie beyond any solver's reach. For example, the current best surface-code resource estimate for factoring a 2048-bit \textsc{rsa} integer assumes a surface code of distance $25$~\cite{gidney2025factor}, which is well beyond any solver's capabilities currently.

\subsection{The Homological View}\label{sec:homological}

\begin{figure}[t]
\centering
\input{figures/toric-grid-ops.tikz}
\caption{$X$-type operators on the $L = 5$ toric code; \textbf{qubits are the edges}. In the shaded boundary, the dotted shared edge cancels.}
\label{fig:toric-grid-ops}
\end{figure}
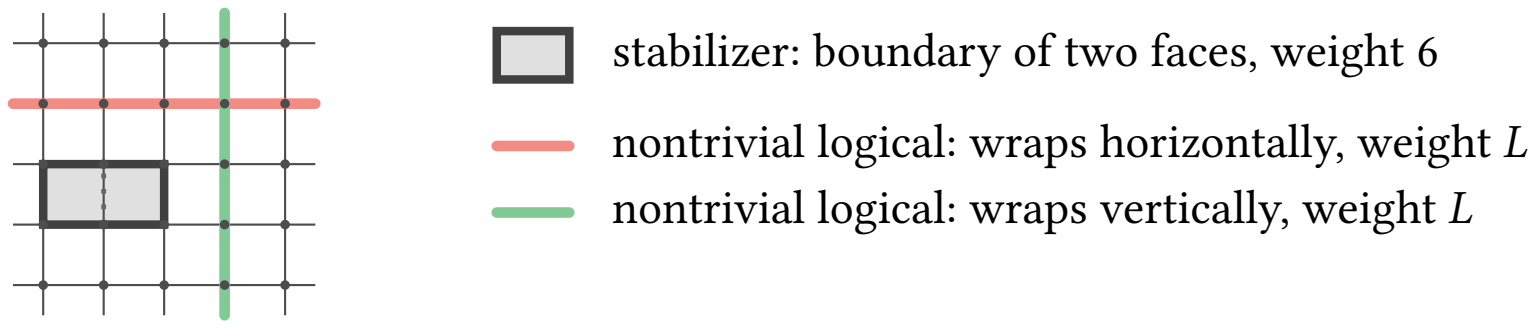

In \cref{sec:distance-def}, we argued that computing distance naïvely doesn't scale to large codes. One approach to address this issue is to use a mathematical representation of a stabilizer code that makes it easier to reason about the nontrivial logical operators of a code.
It turns out that for several code families, one can analyze the nontrivial logical operators of the code \emph{geometrically}~\citep{kitaev2003fault,dennis2002topological}. While the approach is general, throughout this section we will use the toric code as our running example.

\paragraph{Toric Code Construction} Recall that in the toric code from \cref{sec:families}, the qubits are interpreted as the edges of an $L \times L$ periodic grid. The toric code has two kinds of stabilizer generators: $X$-type generators (face) and $Z$-type generators (vertex). Such codes with this clean division into $X$- and $Z$-type stabilizer generators, called Calderbank--Shor--Steane (\textsc{css}) codes~\citep{calderbank1996good,steane1996multiple}, are well-studied and form the majority of the practical codes under serious consideration for the development of fault-tolerant quantum computers.
Indeed, all of the codes in \cref{tab:case-studies} are \textsc{css} codes.

\paragraph{Nontrivial Logical Operators of the Toric Code}  A nontrivial logical operator of the toric code (\cref{def:logical}) is a \emph{loop around the torus}, such as the red and green loops of weight $L$ in \cref{fig:toric-grid-ops}.
To see why, recall that a nontrivial logical operator commutes with all the stabilizer generators but is not itself a stabilizer. Since the toric code is a \textsc{css} code, it suffices to consider $X$-type operators, each of which is a set of edges. Such an operator commutes with every face generator, since the operators commute qubitwise. $X$ anticommutes with $Z$, so an $X$-type operator commutes with the vertex generator at vertex $v$ when it contains an even number of the edges at $v$; an edge set that meets every vertex an even number of times is called a \emph{closed loop}. Finally, multiplying the face generators of two adjacent faces cancels $X$ on their shared edge and leaves $X$ on the six edges around the pair (the shaded patch in \cref{fig:toric-grid-ops}), so every $X$-type stabilizer is the \emph{boundary} of a set of faces. \textbf{Thus, the nontrivial logical operators of the toric code are the closed loops that are not the boundary of any set of faces}, that is, the loops that wrap around the torus.

The distance of the toric code on an $L \times L$ grid is at most $L$, since we have examples of nontrivial logical operators of weight $L$. We now want to argue that every nontrivial logical operator has weight at least $L$. Geometrically, this is clear because every closed loop that doesn't enclose a set of faces must wrap all the way across the torus, which has length $L$.
The language of homology~\citep{dennis2002topological,freedman2001projective,bombin2007homological} makes this geometric intuition precise by viewing the toric lattice through the lens of linear algebra.

\paragraph{Homological View of the Toric Lattice} The homological view algebraically captures the layout of the torus (i.e., the faces, edges, and vertices) as three vector spaces over $\mathbb{F}_2$: we represent a set of faces as a vector with one coordinate per face, and similarly for sets of edges and sets of vertices. The structural relationship between them is captured by two linear maps:
\[
  \mathsf{Faces}
  \xrightarrow{\;\partial_f\;}
  \mathsf{Edges}
  \xrightarrow{\;\partial_e\;}
  \mathsf{Vertices}
\]
$\partial_f$ maps a face to the edges that make up the face, and $\partial_e$ maps an edge to the sum of its endpoints. The three vector spaces and the two linear maps between them are called a \emph{chain complex}.

The following example shows how faces, edges, and vertices are encoded as binary vectors.

\begin{example}[grid box]\label{ex:single-box}
Consider one square with vertices $v_1, \ldots, v_4$, edges $e_1, \ldots, e_4$, and face $f$:
\begin{center}
\begin{tikzpicture}[x=0.8cm, y=0.8cm, every node/.style={font=\footnotesize}]
  \fill[black!10] (0,0) rectangle (1,1);
  \draw[black!70, line width=0.6pt] (0,0) rectangle (1,1);
  \foreach \p/\n/\a in {(0,0)/v_1/below left, (1,0)/v_2/below right, (1,1)/v_3/above right, (0,1)/v_4/above left}
    {\fill \p circle[radius=1.4pt]; \node[anchor=center, \a] at \p {$\n$};}
  \node[below] at (0.5,0) {$e_1$};
  \node[right] at (1,0.5) {$e_2$};
  \node[above] at (0.5,1) {$e_3$};
  \node[left]  at (0,0.5) {$e_4$};
  \node at (0.5,0.5) {$f$};
\end{tikzpicture}
\end{center}
Then $\mathsf{Faces} = \mathbb{F}_2^1$, $\mathsf{Edges} = \mathbb{F}_2^4$, and $\mathsf{Vertices} = \mathbb{F}_2^4$, with one coordinate per face, edge, and vertex. A set of faces, edges, or vertices is represented by the vector with a $1$ at each of its members. The single edge $e_1$ is $(1,0,0,0) \in \mathsf{Edges}$, and the two edges $\{e_1, e_2\}$ are $(1,1,0,0)$. Since $\partial_e$ sends each edge to its endpoints, $\partial_e(e_1) = \{v_1, v_2\}$ and $\partial_e(e_2) = \{v_2, v_3\}$, and by linearity
\[
  \partial_e(1,1,0,0) = (1,1,0,0) + (0,1,1,0) = (1,0,1,0) = \{v_1, v_3\}
\]
The shared vertex $v_2$ is counted twice and cancels, leaving the two loose ends of the path. For all four edges, every vertex is counted twice, so $\partial_e$ gives $0$: the boundary of the face is a closed loop.
\end{example}

\paragraph{Logical Operators Through the Lens of Homology} It turns out that we can represent all nontrivial logical operators in the language of homology, through the \emph{first homology group}. We write $\ker(\partial_e) = \{v \in \mathsf{Edges} : \partial_e v = 0\}$ for the \emph{kernel} of $\partial_e$, the edge sets that $\partial_e$ sends to zero, and $\im(\partial_f) = \{\partial_f f : f \in \mathsf{Faces}\}$ for the \emph{image} of $\partial_f$, the edge sets that are boundaries of sets of faces. Since every boundary is a closed loop, $\im(\partial_f) \subseteq \ker(\partial_e)$.

\begin{definition}[first homology group]\label{def:homology}
The \emph{first homology group} of the chain complex is the quotient $H_1 = \ker(\partial_e) / \im(\partial_f)$, in which two vectors of $\ker(\partial_e)$ are identified when they differ by an element of $\im(\partial_f)$.
\end{definition}

 Intuitively, $\ker(\partial_e)$ gives us all closed loops in the grid, that is, the operators that commute with every stabilizer generator. $\im(\partial_f)$ gives us all boundaries of sets of faces, that is, the stabilizers. So $\ker(\partial_e)/\im(\partial_f)$ is the set of closed loops counted up to boundaries, that is, the logical operators counted up to stabilizers. Its nonzero elements are the nontrivial logical operators. For the toric code, it has four elements: no wrap (zero), horizontal (the red loop in \cref{fig:toric-grid-ops}), vertical (the green loop), and both.

Computing distance amounts to finding the lightest representative of a nonzero element of $H_1$.

\begin{definition}[distance via homology]\label{def:homological-distance}
The \emph{$X$-distance} of the toric code is
\[
  d_X = \min \{\, \mathrm{wt}(v) : v \in \ker(\partial_e) \setminus \im(\partial_f) \,\},
\]
and the $Z$-distance $d_Z$ is defined in the same way on the dual lattice, with faces and vertices exchanged. The distance of the code is $d = \min(d_X, d_Z)$.
\end{definition}

Finally, to prove that the distance of the code is $L$, we need to argue that every element of $\ker(\partial_e) \setminus \im(\partial_f)$ has weight $\geq L$. This can be done by introducing `wrapping invariants' that are nonzero on every nontrivial logical operator in $\ker(\partial_e) \setminus \im(\partial_f)$ and force it to cross every column or every row of the lattice.

\paragraph{Homology in \f} \f formalizes this view with a \lstinline{HomologicalCode} type:
\begin{leancode}
structure HomologicalCode where
  Vertices : Type
  Edges : Type                    -- one qubit each
  Faces : Type
  ∂e : (Edges → ZMod 2) →ₗ[ZMod 2] (Vertices → ZMod 2)
  ∂f : (Faces → ZMod 2) →ₗ[ZMod 2] (Edges → ZMod 2)
  boundary_comp : ∂e.comp ∂f = 0
  ...
\end{leancode}
\begin{sloppypar}
A \lstinline{HomologicalCode} consists of three types corresponding to the three vector spaces, the two boundary maps, and the chain-complex law $\partial_e \circ \partial_f = 0$, which says that every boundary is a closed loop; we omit the finiteness and indexing fields, and the toric code is an instance of this type. \f turns every \lstinline{HomologicalCode} into a \lstinline{StabilizerCode} whose number of logical qubits is $\dim H_1$, and proves that an $X$-type operator is a nontrivial logical operator  when its edge set lies in $\ker(\partial_e) \setminus \im(\partial_f)$, with the symmetric statement for $Z$-type operators.
\end{sloppypar}

\paragraph{Conclusion} Through the language of homology, we were able to exploit the geometric structure of the toric lattice to analyze the nontrivial logical operators of the code. More generally, the language of homology can concisely describe a wide range of practically important \textsc{qec} codes~\citep{bravyi2014homological,breuckmann2021quantum} such as \textsc{bb} codes (\cref{sec:BB}) and mitten codes (\cref{sec:mitten}). In the next section, we give a way to exploit the structural relationship between a pair of codes to reason about the nontrivial logical operators.

\section{Compositional Distance Proofs}\label{sec:distance}

In this section, we introduce the idea of a \emph{\textsc{qec} code combinator}: a construction that builds new codes from existing ones, together with proof rules for the parameters of the codes it builds. \Cref{fig:rules-alt} gives inference rules for two such combinators: the standard \emph{concatenation} rule, and two new rules for \emph{double covers}. The \emph{cover} rule is a lower bound on the distance of a double cover and the \emph{doubling} rule refines it to an exact distance under additional hypotheses. The new rules exploit the structure of the construction to enable distance proofs for many large \textsc{qec} codes.

\begin{figure}[t]
\begin{mathpar}
\inferrule*[right={\hypertarget{rule:concat}{Concat}}]
  {C_{\mathrm{in}} : \nkd{n_1, k_1, d_1} \\ C_{\mathrm{out}} : \nkd{n_2, k_2, d_2}}
  {C_{\mathrm{out}} \circ C_{\mathrm{in}} : \nkd{n_1 n_2,\, k_1 k_2,\, {\geq}\, d_1 d_2}}
\and
\inferrule*[right={\hypertarget{rule:cover}{Cover}}]
  {\Cbase : \nkd{n, k, d} \\ \Ccov \text{ is a double cover of } \Cbase \\ d' \leq 2d \text{ as in \cref{thm:cover-bound}}}
  {\Ccov : \nkd{2n,\, k',\, {\geq}\, d'}}
\and
\inferrule*[right={\hypertarget{rule:double}{Double}}]
  {\Cbase : \nkd{n, k, d} \\ \Ccov \text{ is a double cover of } \Cbase \\ \text{conditions of \cref{cor:doubling}}}
  {\Ccov : \nkd{2n,\, k',\, 2d}}
\end{mathpar}
\caption{Code combinators. \ruleConcat encodes each qubit of an outer code with an inner code. \ruleCover glues two copies of a base code $\Cbase$ into a double cover $\Ccov$ and bounds its distance below by a quantity $d^{\prime} \leq 2d$ (\cref{sec:doubling}, \cref{thm:cover-bound}). \ruleDouble refines the bound to $2d$ under additional hypotheses. In both, $k'$ of $\Ccov$ satisfies $0 \leq k' \leq 2k$ and is determined by $k$ and the structure of the double cover (\cref{app:double-cover}).}
\label{fig:rules-alt}
\end{figure}

\subsection{Code Concatenation}\label{sec:concat}

Concatenation is the standard way to build a code of higher distance from two smaller codes~\citep{knill1996concatenated,gottesman1997stabilizer}. Given an \emph{inner} code $C_{\mathrm{in}}$ and an \emph{outer} code $C_{\mathrm{out}}$, we replace each physical qubit of the outer code by a block of the inner code, so that the logical qubits of the inner blocks serve as the physical qubits of the outer code.

\ruleConcat in \cref{fig:rules-alt}---which is formalized in \f---gives the parameters of the new code. Intuitively, a nontrivial logical operator of the new code is a nontrivial logical operator of the outer code, of weight at least $d_2$, in which each qubit is replaced by a nontrivial logical operator of the inner code, of weight at least $d_1$; hence its weight is at least $d_1 d_2$.

\paragraph{Example} Concatenating the $\nkd{7, 1, 3}$ Steane code (\cref{fig:color-patches}) with itself gives
\[
  \inferrule*[right={\ruleConcat}]
    {\mathrm{Steane} : \nkd{7, 1, 3} \\ \mathrm{Steane} : \nkd{7, 1, 3}}
    {\mathrm{Steane} \circ \mathrm{Steane} : \nkd{49, 1, 9}}
\]
that is, $7 \cdot 7 = 49$ qubits encoding $1 \cdot 1 = 1$ logical qubit with distance $3 \cdot 3 = 9$.

\subsection{The Double Cover Combinator}\label{sec:doubling}

Concatenation can build codes of any distance, but most codes of practical interest, such as those in \cref{sec:case-studies}, are not concatenated from smaller codes. We now develop the \emph{double cover combinator}: a construction, the \emph{double cover}, together with two proof rules, \ruleCover and \ruleDouble, for the distance of the code it builds.

The construction takes a \emph{base} code $\Cbase$ and glues two copies of it into a \emph{double cover} code $\Ccov$ on twice as many qubits. It generalizes a construction of \citet{symons2025sequences} from bivariate bicycle codes to arbitrary stabilizer codes; for example, the $\nkd{144, 12, 12}$ gross code is a double cover of a $\nkd{72, 12, 6}$ code. The proof rules then relate the distance of $\Ccov$ to that of $\Cbase$: we reason about the logical operators of the smaller and simpler code, and \emph{scale} that reasoning to the larger one.

We provide a semi-formal description of the two rules and the construction. See \cref{app:double-cover} for a comprehensive treatment.

\begin{figure}[t!]
\centering
\begin{tikzpicture}[x=0.55cm, y=0.55cm,
  qubit/.style={circle, fill=black, inner sep=1.5pt},
  lab/.style={font=\small},
  chk/.style={rounded corners=4pt, draw=red!70!black, fill=red!12, thick},
  pair/.style={gray!45, thin}]
  \node[lab] at (3.5, 1.6) {base code $\Cbase$};
  \draw[chk] (1.55, -0.45) rectangle (4.45, 0.45);
  \node[lab, red!70!black] at (3, 0.95) {$b$};
  \foreach \j in {1,...,6} {
    \node[qubit] (b\j) at (\j, 0) {};
    \node[lab] at (\j, -1.05) {$\j$};
  }
  \draw[->, thick] (3.65cm, 0) -- (4.85cm, 0) node[lab, midway, above] {lift};
  \begin{scope}[xshift=4.7cm]
    \node[lab] at (3.5, 2.2) {double cover $\Ccov$};
    \foreach \j in {1,...,6} { \draw[pair] (\j, 0.8) -- (\j, -0.8); }
    \draw[chk] (1.55, 0.35) rectangle (3.45, 1.25);
    \draw[chk] (3.55, -1.25) rectangle (4.45, -0.35);
    \node[lab, red!70!black] at (2.5, 1.7) {$a$};
    \node[lab, red!70!black] at (4, -1.7) {$a \cdot b$};
    \foreach \j in {1,...,6} {
      \node[qubit] at (\j, 0.8) {};
      \node[qubit] at (\j, -0.8) {};
    }
    \node[lab, anchor=west] at (6.5, 0.8) {sheet $0$};
    \node[lab, anchor=west] at (6.5, -0.8) {sheet $1$};
    \draw[<->, thick] (8.9, 0.55) to[bend left=60] node[lab, right] {$\sigma$} (8.9, -0.55);
  \end{scope}
\end{tikzpicture}
\caption{Doubling the qubits of a base code $\Cbase$. Left: six base qubits and a stabilizer generator $b$ on qubits $2$--$4$. Right: the two sheets of a double cover $\Ccov$; the two copies of each base qubit sit in one column, joined by a gray line. Swap $\sigma$ exchanges the sheets. The stabilizer generator $b$ lifts to the stabilizer $s = (a, a \cdot b)$ of $\Ccov$, which acts as $a$ on qubits $2$ and $3$ of sheet $0$ and as $a \cdot b$ on qubit $4$ of sheet $1$; folding $s$ recovers $b$.}
\label{fig:sheets}
\end{figure}
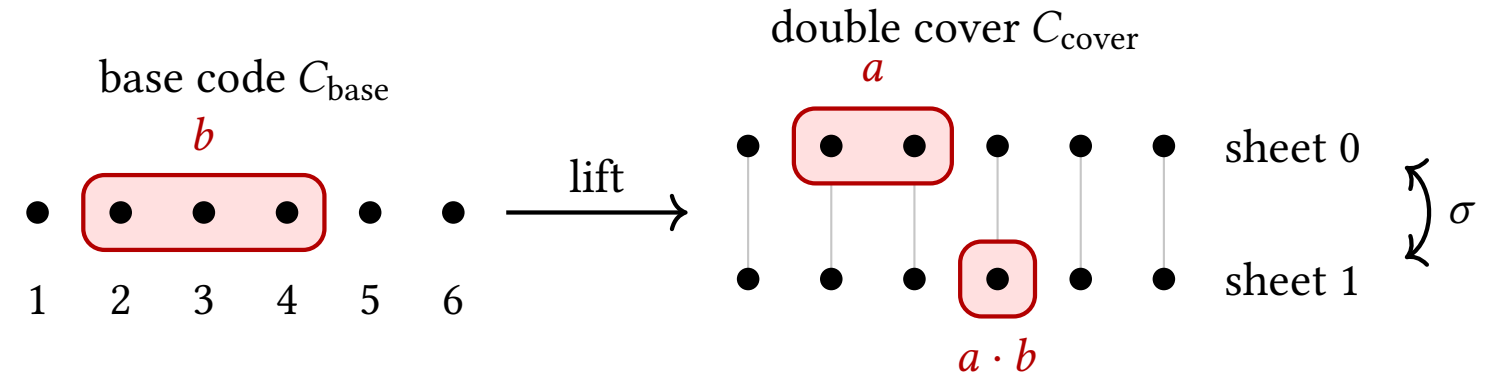

\paragraph{Double Covers}
Let $\Cbase$ be a code on $n$ qubits. A double cover $\Ccov$ is defined over two copies of these qubits, which we call \emph{sheet $0$} and \emph{sheet $1$} (\cref{fig:sheets}). We write an operator on the $2n$ qubits as a pair $P = (u, u')$ of operators on the base qubits, $u$ on sheet $0$ and $u'$ on sheet $1$, and write $P^\sigma = (u', u)$ for its \emph{mirror image}. We define two maps that relate the operators of $\Ccov$ and $\Cbase$.

\begin{definition}[fold and double]\label{def:fold-double-main}
To \emph{fold} an operator of $\Ccov$, we lay one sheet on top of the other and multiply: $\operatorname{fold}(u, u') = u \cdot u'$. To \emph{double} an operator of $\Cbase$, we copy it onto both sheets: $\operatorname{double}(u) = (u, u)$.
\end{definition}

\noindent Observe that doubling $u$ doubles its weight, and that folding $(u,u')$ yields an operator of weight at most that of $(u,u')$: $\operatorname{fold}$ is a weight non-increasing map.

\paragraph{Building a Double Cover}
The main idea of the double cover combinator is that we can take the stabilizer generators of $\Cbase$ on $n$ qubits and \emph{lift} them to operators on $2n$ qubits to obtain the stabilizer generators of $\Ccov$~\citep{guemard2024lifts,symons2025sequences}. Consider the construction on the six-qubit base code of \cref{fig:sheets}. Concretely, consider the stabilizer generator $b$, which acts on qubits $2$, $3$, and $4$. To lift $b$ to $\Ccov$, we decide which part of it goes on sheet $0$. Here we put its part $a$ on qubits $2$ and $3$ on sheet $0$, and the rest of $b$, which is $a \cdot b$ on qubit $4$, on sheet $1$. This gives the lifted generator $s = (a, a \cdot b)$. Note that folding $s$ recovers $b$, since the two copies of $a$ cancel: $\operatorname{fold}(s) = a \cdot a \cdot b = b$. For $\Ccov$, we also add the mirror image $s^\sigma = (a \cdot b, a)$, which places the same two pieces on the opposite sheets. We do this for every generator of $\Cbase$. The lifts together with their mirror images generate the stabilizer group of $\Ccov$. In summary, we define a double cover as follows:

\begin{definition}
  A stabilizer code $\Ccov$ on $2n$ qubits is a \emph{double cover} of a stabilizer code $\Cbase$ on $n$ qubits if its stabilizers are closed under mirroring, and folding them yields the stabilizers of $\Cbase$.
\end{definition}

The choice of splits decides how the two copies are connected. If we keep every generator whole on sheet $0$, then $\Ccov$ is just two disjoint copies of $\Cbase$. In this case, the distance of $\Ccov$ is the same as that of $\Cbase$ since the nontrivial logical operators of $\Cbase$ are also nontrivial logical operators of $\Ccov$. Thus the cover code under this trivial split is not interesting. Instead, if we split some generators across the two sheets, as $b$ is in \cref{fig:sheets}, they straddle the two sheets and glue the copies together along a \emph{seam}. This complicates the structure of the logical operators of $\Ccov$ and can sometimes lead to codes with higher distance. Next, we discuss our proof rules that reason about the relationship between the logical operators of $\Cbase$ and $\Ccov$.

\paragraph{Distance of Double Covers} We propose two rules, \ruleCover and \ruleDouble, that reason about the distance of $\Ccov$ through its structural relationship with $\Cbase$:
\begin{itemize}
  \item \ruleCover gives a lower bound $d' \leq 2d$ on the distance of $\Ccov$. That is, if $\Cbase$ is an $\nkd{n, k, d}$ code, then its double cover $\Ccov$ is a $\nkd{2n, k', {\geq}\, d'}$ code.
  \item \ruleDouble gives conditions under which the distance of $\Ccov$ is exactly $2d$.
\end{itemize}
Intuitively, folding a nontrivial logical operator of $\Ccov$ yields either the identity, in which case the operator is the double of a nontrivial logical operator of $\Cbase$ and has weight at least $2d$, or a non-identity operator of $\Cbase$, whose weight bounds the operator's weight from below. \ruleDouble's conditions, which can be checked on $\Cbase$ alone (\cref{app:double-cover}), ensure that the second case also yields weight at least $2d$. In both rules, the number of logical qubits $k'$ of $\Ccov$ satisfies $0 \leq k' \leq 2k$ and is determined by $k$ and the structure of the double cover (\cref{app:double-cover}).

\section{Implementation and Case Studies}\label{sec:case-studies}

In this section, we (1) discuss the implementation of \f (\cref{sec:implementation}), and (2) present case studies that showcase the generality and power of our framework (\cref{sec:BB} onward).

\subsection{Implementing \f}\label{sec:implementation}
We first built the mathematical foundations of \f, such as Pauli groups and Hilbert spaces, and the key abstractions for stabilizer codes. These were written manually, building on Mathlib~\citep{mathlib2020}, and form our \emph{trusted specification layer}. For formalizing and verifying the more recent \textsc{qec} codes, like the \textsc{bb} codes, we extensively used \textsc{ai} agents.

In order to trust that a \textsc{qec} code has been formalized correctly, one only has to check that the Lean stabilizer generators with which the \lstinline|StabilizerCode| specification is instantiated match the code's informal description in the literature. This means that we do not need to explicitly trust the complex proofs \textsc{ai} agents may generate to construct a valid \textsc{qec} code or prove a code's distance once the specifications have been accepted.

The foundations of \f, consisting of the mathematical foundations, code abstractions, and combinators of \cref{fig:architecture}, comprise about $25{,}000$ lines of Lean. The formalized codes and their proofs comprise about $32{,}000$ further lines, along with about $312{,}000$ lines of machine-generated certificates that Lean's kernel checks, which belong primarily to the distance proofs of the $\nkd{360, 12, 24}$ \textsc{bb} code and the mitten code. Every declaration in \f depends only on Lean's three standard axioms.

\subsection{Case Study: Bivariate Bicycle Codes}\label{sec:BB}

Quantum computing architectures based on superconducting qubits are one of the leading contenders for fault-tolerant quantum computers. Good quantum error-correcting codes for such architectures need to support the limited connectivity between the physical qubits while also having a high rate and high distance. The \emph{bivariate bicycle} (\textsc{bb}) code family, introduced by \citet{bravyi2024high}, is a leading family of \textsc{qec} codes for fault-tolerant quantum hardware based on superconducting qubits. A subfamily identified by \citet{bravyi2024high} meets the connectivity constraints imposed by the architecture and consists of instances like the so-called \emph{gross} code $\nkd{144, 12, 12}$ and the $\nkd{360, 12, 24}$ \textsc{bb} code, which have high rate and high distance.

We formalize the construction of the entire \textsc{bb} code family as a parametrized \lstinline|StabilizerCode n k| and additionally formalize the distance for the gross $\nkd{144, 12, 12}$ \textsc{bb} code and then the $\nkd{360, 12, 24}$ \textsc{bb} code using novel structural arguments as discussed in \cref{sec:distance}.

\paragraph{Constructing \textsc{bb} Codes}
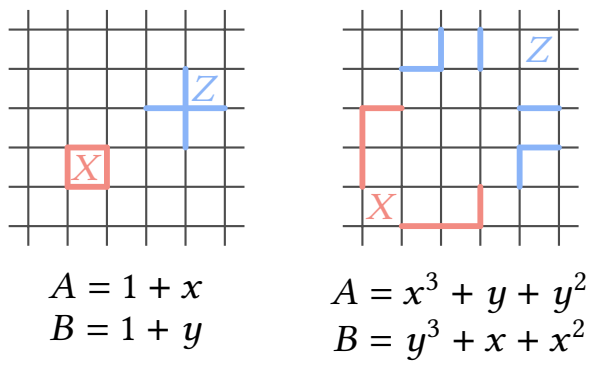
\begin{wrapfigure}{r}{0.3\linewidth}
  \vspace{-1em}
  \centering
  \input{figures/bb-stencils.tex}
  \caption{One $X$ and one $Z$ generator of two \textsc{bb} codes on a $6 \times 6$ torus; qubits are the edges. Left: toric code. Right:  $\nkd{72, 12, 6}$~\citep{bravyi2024high}.}
  \label{fig:bb-stencils}
  \vspace{-1em}
\end{wrapfigure}

Recall the toric code, which we illustrated in \cref{sec:families}.
The qubits are laid out on the surface of a torus, which is flattened into a grid, with the stabilizer generators acting on neighboring qubits.
\textsc{bb} codes generalize this construction into an entire family of codes.
Interestingly, a \textsc{bb} code can be defined using two polynomials, $A$ and $B$, along with an underlying abelian group $\mathbb{Z}_\ell \times \mathbb{Z}_m$.
By changing the definitions of $A$ and $B$, we get different \textsc{bb} codes.
Researchers have been searching for polynomials that induce desirable qualities in the code.

\Cref{fig:bb-lean} shows the Lean definition of a \textsc{bb} code. The constructor \lstinline|bbStabilizerCode| takes two polynomials (and torus dimensions $\ell$ and $m$) and generates a \lstinline|Code[[n, k]]|.
We show a number of definitions of \textsc{bb} codes, including the ones on \textsc{ibm}'s roadmap, e.g., the gross code.
We also show the toric code family defined in this notation. \Cref{fig:bb-stencils} visualizes the stabilizer generators of two of these codes: the toric code and \lstinline|bb72|.

\begin{figure}[t]
\begin{leancode}
noncomputable def bbStabilizerCode (ℓ m : ℕ) [NeZero ℓ] [NeZero m]
    (A B : Poly ℓ m) : Code[[2 * (ℓ * m), 2 * dim (ker A ⊓ ker B)]]

-- Examples of BB codes from Bravyi et al.; bb144 is the gross code
def bb72  := bbStabilizerCode  6  6 poly[x^3 + y + y^2]   poly[y^3 + x + x^2]
def bb144 := bbStabilizerCode 12  6 poly[x^3 + y + y^2]   poly[y^3 + x + x^2]
def bb288 := bbStabilizerCode 12 12 poly[x^3 + y^2 + y^7] poly[y^3 + x + x^2]
def bb360 := bbStabilizerCode 30  6 poly[x^9 + y + y^2]   poly[y^3 + x^25 + x^26]

-- The toric code family, up to a relabeling of qubits
def toricBB (L : ℕ) [NeZero L] := bbStabilizerCode L L poly[1 + x] poly[1 + y]
\end{leancode}
\caption{Bivariate bicycle codes in \f. The constructor \lstinline{bbStabilizerCode} takes the torus dimensions $\ell$, $m$ and the polynomials $A$, $B$, and returns a stabilizer code on $2\ell m$ qubits. Below it, four \textsc{bb} codes of \citet{bravyi2024high} and the toric code family}
\label{fig:bb-lean}
\end{figure}

\paragraph{Formalizing the $\nkd{360, 12, 24}$ \textsc{bb} Code}
No closed-form distance formula has been found so far for \textsc{bb} codes. As a result, the primary method to compute the distance of a \textsc{bb} code is to use simulation or a constraint solver. Unfortunately,  these solvers do not scale to \textsc{bb} codes with distances $\geq 24$, and so only upper bounds for distance are known for many \textsc{bb} codes---which do not provide any conclusive guarantee on the power of the code to detect and correct errors. We prove the distances of the two \textsc{bb} codes with the rules of \cref{sec:distance}, in different ways. The gross code is a double cover of a $\nkd{72, 12, 6}$ code that meets the hypotheses of \ruleDouble, which a kernel-checked certificate on the 72-qubit base discharges. The $\nkd{360, 12, 24}$ code, in contrast, does not meet them. Its proof follows the case analysis of \ruleCover down a tower of double covers on $360$, $180$, and $90$ qubits. \Cref{app:bb360} outlines the proof. We argue that our foundations in \f can be used to automatically discover and verify novel \textsc{bb} codes with high rate and high distance via large-scale agentic search.

\subsection{Case Study: Toric and Rotated Surface Codes}\label{sec:toric-case-study}
Kitaev's toric code family~\citep{kitaev2003fault} was the first code family to be discovered whose distance grows with lattice size while every stabilizer generator acts only on neighboring qubits, as discussed in \cref{sec:BB}. Its periodic boundaries, however, cannot be laid out on a planar chip with nearest-neighbor connectivity, which is what hardware architectures like superconducting qubits provide. The (rotated) \emph{surface codes} are the planar variants of the toric code obtained by cutting the torus open and introducing boundaries~\citep{dennis2002topological}; they encode a single logical qubit and are local on a two-dimensional grid.  For example, Google used the rotated surface code with code parameters $\nkd{L^2, 1, L}$ to demonstrate that the logical error rate (i.e., error rate after applying error correction) falls with increasing code distance~\citep{googleqai2023suppressing,googleqai2025below}. We've already discussed the construction of the toric code in \cref{sec:families,sec:homological}. We use the homological distance argument and formalize the entire toric code family as a \lstinline|StabilizerCodeWithDistance|:

\begin{leancode}
theorem toricCodeN_distance_eq_L (L : ℕ) [Fact (2 ≤ L)] :
    HasCodeDistance (toricStabilizerCode L) L

noncomputable def toricStabilizerCodeWithDistance (L : ℕ) [Fact (2 ≤ L)] :
    Code[[2 * L * L, 2, L]] where
  toStabilizerCode := toricStabilizerCode L
  hasDistance      := toricCodeN_distance_eq_L L
\end{leancode}

\noindent The theorem is the endpoint of the homological argument (\cref{sec:homological}). To our knowledge, this is the first machine-checked formalization of the toric code family---or of any code family, for that matter.

\begin{wrapfigure}{r}{0.32\linewidth}
\vspace{-1em}
\centering
\begin{tikzpicture}[x=0.32cm, y=0.32cm]
  \surfacepatch{5}
  \begin{scope}[shift={(5.6,0.2)},
                q/.style={fill=black!85}, lbl/.style={font=\scriptsize, anchor=west}]
    \filldraw[chk, fill=xstab] (0,2.6) rectangle ++(1,1);
    \node[lbl] at (1.2,3.1) {$X^{\otimes 4}$};
    \filldraw[chk, fill=zstab] (0,1.2) rectangle ++(1,1);
    \node[lbl] at (1.2,1.7) {$Z^{\otimes 4}$};
    \filldraw[chk, fill=xstab] (0,0) arc[start angle=180, end angle=0, radius=0.5] -- cycle;
    \node[lbl] at (1.2,0.25) {$X^{\otimes 2}$/$Z^{\otimes 2}$};
  \end{scope}
\end{tikzpicture}
\caption{Rotated surface code, $L = 5$; \textbf{qubits are the vertices}.}
\label{fig:surface-patches}
\vspace{-1em}
\end{wrapfigure}
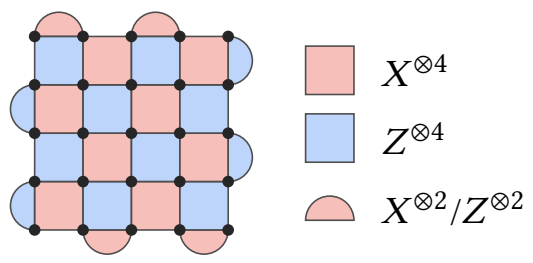
\paragraph{Rotated Surface Code}
The rotated surface code is preferred over the unrotated planar surface code for practical implementations because it reaches the same distance with roughly half the physical qubits, while its stabilizer generators still have weight at most four and still act on nearest neighbors in a two-dimensional grid~\citep{bombin2007optimal,tomita2014low}. As \cref{fig:surface-patches} shows, the data qubits sit on the vertices of an $L \times L$ grid, with weight-four $X$ and $Z$ generators alternating in a checkerboard on the interior faces and weight-two generators along the boundary. \f formalizes this family by the same homological route as the toric code, packaging it as a \lstinline|Code[[L * L, 1, L]]| for odd $L \geq 3$.

\subsection{Case Study: Mitten Codes}\label{sec:mitten} While \textsc{bb} codes and the rotated surface codes are practically important because of their low-weight generators, local or nearly local two-dimensional layouts, and families whose distance grows, their encoding rate $\frac{k}{n}$ is relatively poor. This means that large-scale fault-tolerant architectures using these codes will incur high physical qubit overhead. For instance, the rotated surface code with distance 12 has rate $\frac{1}{144}$ and the gross code has rate $\frac{1}{12}$. \citet{bhardwaj2026mitten} introduce the \emph{mitten code} family, which has a constant rate of $\frac{1}{5}$, but sacrifices the low-weight stabilizer generators and locality. The resulting non-local stabilizer generators are a natural fit for neutral-atom architectures, where qubit connectivity is not fixed by the chip layout as in superconducting computers~\citep{bhardwaj2026mitten}.

Mitten and \textsc{bb} codes are both instances of the \emph{lifted product} construction of \citet{panteleev2021degenerate,panteleev2022quantum}, which lifts the stabilizer generators of a small base code through the algebra of a group acting on the qubits: the abelian group $\mathbb{Z}_\ell \times \mathbb{Z}_m$ for \textsc{bb} codes (\cref{sec:BB}) and a non-abelian group for mitten codes. In \f, we have a generic construction of mitten codes as a \lstinline|HomologicalCode| and prove all three code parameters of the $\nkd{150, 30, 10}$ mitten code. Notably, this instance lies outside every family discussed so far, yet its distance proof goes through the same homological bridge as the toric code in \cref{sec:homological}, with the classification of low-weight cycles discharged by kernel-checked certificates.

\section{Related Work}

\paragraph{Quantum Error Correction}
Recent experiments demonstrate error suppression with \textsc{qec} codes on several architectures~\citep{googleqai2023suppressing,googleqai2025below,bluvstein2026fault,paetznick2024demonstration,putterman2025hardware} and logical operations on error-corrected qubits~\citep{bluvstein2024logical,reichardt2024fault,lacroix2025scaling,salesrodriguez2025experimental}. On the theory side, q\textsc{ldpc} codes and code families~\citep{tillich2014quantum,kovalev2013quantum,panteleev2021degenerate,breuckmann2021balanced,panteleev2022asymptotically,bravyi2024high,bhardwaj2026mitten} are leading candidates for fault-tolerant quantum computers. Several works show how to perform logical operations on these codes~\citep{cohen2022low,cross2024improved,williamson2026low,eberhardt2025logical,xu2024constant,yoder2025tour,he2025extractors}. Finally, recent resource estimates for Shor's algorithm target \textsc{rsa}-2048 factoring and elliptic-curve discrete logarithms~\citep{gidney2025factor,webster2026pinnacle,cain2026shor,babbush2026securing}. All of these rely on code parameters, which \f formally verifies.
\paragraph{Code Distance Computation}
Computing the distance of a \textsc{qec} code is \textsc{np}-hard~\citep{kapshikar2023hardness}. Several approaches use randomized and heuristic upper bounds~\citep{pryadko2022qdistrnd,bravyi2024high,webster2026distance}, but these cannot rule out a lighter logical operator. Exact methods use enumeration~\citep{dumer2017distance,hernando2026fast} or \textsc{sat} and \textsc{ilp} solvers~\citep{bravyi2024high,chen2026sat,webster2026distance}, but do not scale to large codes or to code families. For specific families, like toric and \textsc{bb} codes, several structural results reason about distance~\citep{kitaev2003fault,dennis2002topological,lin2024quantum,symons2025sequences,guemard2024lifts}. \f contributes a trusted Lean specification of distance, against which researchers and \textsc{ai} tools can verify new structural reductions, and two such reductions (\cref{sec:concat,sec:doubling}).
Our double covers generalize the two-fold cover codes of \citet{symons2025sequences}, who prove the bounds $d \leq d_h \leq hd$ on the distance $d_h$ of an $h$-fold cover of a \textsc{bb} code only for odd $h$; \ruleCover and \ruleDouble treat $h = 2$. Closely related to our work, \citet{ehatamm2026end} also formalize stabilizer codes in Lean; they verify a reduction from distance to \textsc{sat} and certify distance lower bounds for individual codes by solver search. In contrast, \f's structural reductions derive the distance of a large code from that of a smaller one, which lets us prove distances out of reach for solvers, such as $d = 24$ for the $\nkd{360, 12, 24}$ \textsc{bb} code, and the distance of entire code families. Concurrently, QECert~\citep{he2026qecert} formalizes asymptotic \textsc{qec} results in Lean, such as parameter theorems for product constructions of q\textsc{ldpc} codes and locality bounds; it complements \f, which verifies parameters of concrete codes and code families.

\paragraph{Code Discovery}
While most \textsc{qec} code discovery is undertaken by \textsc{qec} researchers, there is a growing body of work discovering ways to automate the process. For instance, reinforcement learning and related optimization techniques were used by several groups to discover codes given hardware and performance desiderata~\citep{olle2024simultaneous,su2025discovery,he2025discovering,sella2026designing,freire2025optimizing,webster2025engineering,chengyu2026bayesian}. More recently, there have been several frameworks to create \textsc{llm}-driven pipelines to discover codes~\citep{cruzbenito2026evolutionary,liu2026large,yan2026omniqec}. Our work complements these approaches by providing formal specifications for codes and the parameters to optimize over, against which discovered codes can be verified. Beyond \textsc{qec}, \textsc{ai}-for-quantum discoveries are increasingly common~\citep{ruiz2025quantum,shirobokov2026discovering}, signaling a shift in how quantum computing development is done. We argue that formal frameworks of fundamental quantum abstractions, like \f, are an effective way to trust and scale these discoveries.

\paragraph{Verifying Fault-Tolerant Circuits}
From the Programming Languages community, there are several techniques that verify circuit-level implementations of \textsc{qec}. Program logics~\citep{wu2021qecv,huang2025efficient} and symbolic execution~\citep{fang2024symbolic,chen2025verifying} take a fixed code and its error-correction program, and check that the program restores the encoded state for every error in a bounded set, or that a logical operation is fault-tolerant. In parallel, \textsc{zx}-calculus techniques~\citep{rodatz2025fault,rusch2025completeness,bombin2024unifying} transform an idealized fault-tolerant circuit into an implementable one using only rewrites that introduce no new undetectable faults, so that the resulting circuit is fault-tolerant by construction. All of these reason about the implementation of a given code rather than the mathematical construction of the code itself. \f~lays the formal foundation that can be extended to circuit-level verification efforts.

\section{Conclusion}

This paper laid a formal foundation for \textsc{qec}, with the goal of eventually building end-to-end, fully verified quantum compilers that go from high-level applications down to physical qubits, as well as a basis for discovering and exchanging new \textsc{qec} codes.
We see a number of next steps.
First, we plan to automatically construct syndrome-extraction circuits for a given \textsc{qec} code, along with a noise model of the target architecture and its connectivity constraints.
Second, we plan to build logical operations for codes.
Here, our approach is to formalize standard transversal operations as well as something akin to generalized lattice surgery, which applies to all codes discussed above.
Together, these will form the basis for fully verified fault-tolerant quantum computation.
Finally, distance alone does not tell us how a code family behaves under realistic noise; formalizing noise models and the \emph{thresholds} of code families is a natural next step.

\begin{acks}
We thank the members of the \href{https://madpl.cs.wisc.edu/}{MadPL group} for their feedback. This work was supported by the National Science Foundation under grant CCF-2446711.
\end{acks}

\bibliographystyle{ACM-Reference-Format}
\bibliography{references}
\newpage
\appendix
\crefalias{section}{appendix}
\crefalias{subsection}{subappendix}
\section{Double Covers in Full}\label{app:double-cover}

This appendix gives the full definitions behind the double cover combinator of \cref{sec:doubling} and states its two proof rules, \ruleCover (\cref{thm:cover-bound}) and \ruleDouble (\cref{cor:doubling}), precisely. For brevity, we write $B$ for the base code $\Cbase$ and $D$ for the double cover $\Ccov$.

\paragraph{Setup} Consider a stabilizer code $B$ defined on $n$ qubits with stabilizer group $G_B$. We will call this the \emph{base code}. The nontrivial logical operators of the base code $B$ are given by $C(G_B) \setminus G_B$, up to phase, where $C(G)$ is the centralizer of $G$ (Definition~\ref{def:logical}). Our double cover combinator constructs a new stabilizer code $D$ defined on $2n$ qubits with stabilizer group denoted by $G_D$. In order to define our code combinator, we need a way to relate the qubit indices of $D$ to the qubit indices of $B$.

\paragraph{Operators up to Phase} Folding respects the group structure only once phases are discarded (with phases, $\operatorname{fold}(P \cdot Q)$ and $\operatorname{fold}(P) \cdot \operatorname{fold}(Q)$ can differ by a sign), so in this appendix we work in the \emph{phase-free Pauli group}
\[
  \overline{P}_n := P_n / \{\pm I, \pm iI\},
\]
whose elements are the tensor products of $I$, $X$, $Y$, $Z$, multiplied qubit by qubit with the phase discarded. Two elements of $\overline{P}_n$ \emph{commute} if their representatives in $P_n$ do, that is, if their symplectic vectors $(x \mid z)$ and $(x' \mid z')$ (\cref{sec:symplectic}) satisfy $x \cdot z' + z \cdot x' = 0$ in $\mathbb{F}_2$. Like the weight, this does not depend on the phase.

From here on, an \emph{operator} is an element of $\overline{P}_n$, and for a stabilizer group $G$ we write $G$ and $C(G)$ for the images of $G$ and of its centralizer in $\overline{P}_n$. Since $-I \notin G$, distinct stabilizers have distinct images, and the nontrivial logical operators of Definition~\ref{def:logical} are exactly the elements of $C(G) \setminus G$, so the distance is unchanged. \f defines fold and double on phase-free operators in the same way.

\begin{definition}[two sheets, pairs, swap]\label{def:sheets}
Take two copies of the $n$ qubits of $B$, which we call \emph{sheet $0$} and \emph{sheet $1$}, and label the resulting $2n$ qubits by $(j, 0)$ and $(j, 1)$ for $j \in \{1, \ldots, n\}$, according to their sheet. The two copies $(j, 0)$ and $(j, 1)$ of a base qubit $j$ form a \emph{pair}. Since the $2n$ qubits are two copies of the $n$ base qubits, an operator $P$ on the $2n$ qubits can be viewed as a pair of operators on the base qubits: we write $P = (u, u')$, where $u$ acts on the $n$ qubits of sheet 0 and $u'$ acts on the $n$ qubits of sheet 1. Finally, the \emph{swap} function $\sigma$ is the permutation of the $2n$ qubits that exchanges $(j, 0)$ and $(j, 1)$ for every $j$. We write $P^\sigma$ for $P$ with its qubits permuted by $\sigma$, so that $P^\sigma = (u', u)$.
\end{definition}

\Cref{fig:sheets} illustrates the construction on a six-qubit base code.

The two sheets of $n$ qubits of $B$ will be precisely the $2n$ qubits of the double cover $D$. Next, we define functions that relate operators of $D$ with the operators of $B$ in a way that preserves the structure of logical operators.

\begin{definition}[fold and double]\label{def:fold-double}
Define the $\operatorname{fold} : \overline{P}_{2n} \to \overline{P}_n$ and $\operatorname{double}: \overline{P}_n \to \overline{P}_{2n}$ functions on phase-free operators by
\[
  \operatorname{fold}(u, u') := u \cdot u',
  \qquad
  \operatorname{double}(u) := (u, u),
\]
where the product $u \cdot u'$ is taken in $\overline{P}_n$.
\end{definition}

One can check that both maps are group homomorphisms of the phase-free Pauli groups, and for all operators $u \in \overline{P}_n$ and $P \in \overline{P}_{2n}$ they satisfy

\begin{enumerate}
  \item $\operatorname{fold}(\operatorname{double}(u)) = I$
  \item $\operatorname{double}(\operatorname{fold}(P)) = P \cdot P^\sigma$
  \item $\mathrm{wt}(\operatorname{double}(u)) = 2\,\mathrm{wt}(u)$
  \item $\mathrm{wt}(\operatorname{fold}(P)) \leq \mathrm{wt}(P)$
  \item $\operatorname{double}(u)$ commutes with $P$ if and only if $u$ commutes with $\operatorname{fold}(P)$
\end{enumerate}

\paragraph{Constructing the Double Cover $D$ of $B$} We want the stabilizer generators of $D$ to be copies of the stabilizer generators of $B$, except that the qubits in the support of each copy may lie on either sheet. Concretely, fix generators $b_1, \ldots, b_r$ of $G_B$. For each $b_i$, decide which part of it goes to sheet $0$: choose an operator $a_i$ on the base qubits, and lift $b_i$ to the $2n$-qubit operator
\[
  s_i := (a_i,\; a_i \cdot b_i),
\]
which acts as $a_i$ on sheet $0$ and as the remaining part $a_i \cdot b_i$ on sheet $1$. Folding $s_i$ recovers $b_i$, since $\operatorname{fold}(s_i) = a_i \cdot a_i \cdot b_i = b_i$. In the simplest case $a_i$ is $b_i$ restricted to some of its qubits, so that $s_i$ merely distributes the qubits of $b_i$ over the two sheets, as in \cref{fig:sheets}; in general $a_i$ may also act outside the support of $b_i$.

The stabilizer generators of $D$ are the lifts $s_1, \ldots, s_r$ together with their swaps $s_1^\sigma, \ldots, s_r^\sigma$, which induces a symmetry between sheets. For these $2r$ operators to generate a stabilizer group they must pairwise commute, which holds  when
\[
  a_i \text{ commutes with } b_j \iff a_j \text{ commutes with } b_i \qquad \text{for all } i, j.
\]
Under this condition, an independent subset of them, taken with phase $+1$, is a presentation in the sense of \cref{sec:stabilizer}, and we let $G_D$ be the stabilizer group it generates.

The choice $a_i = b_i$ places every generator on sheet $0$ and its swap on sheet $1$, so $D$ is simply two disjoint copies of $B$. Any other choice splits some generators across the two sheets, and we picture $D$ as two copies of $B$ glued along a \emph{seam} that the split generators cross (\cref{fig:sheets}). Under this construction of the double cover $D$ of $B$, one can show that swap maps stabilizers of $D$ to stabilizers of $D$, and the folds of the stabilizers of $D$ are the stabilizers of $B$. Since our proof rule only requires these two properties of the double cover construction, we use them as a definition of a double cover.

\begin{definition}[double cover]\label{def:double-cover}
We say that a stabilizer code $D$ on the two sheets of $B$ is a \emph{double cover} of $B$ if
\begin{enumerate}
  \item the swap preserves its stabilizers, i.e., $g^\sigma \in G_D$ for every $g \in G_D$; and
  \item the folds of its stabilizers are the stabilizers of $B$, i.e., $\operatorname{fold}(G_D) = G_B$.
\end{enumerate}
\end{definition}

Importantly, fold and double preserve the logical structure of operators in the double cover $D$. In particular, one can show that $\operatorname{fold}$ maps logical operators of $D$ to logical operators of $B$, and $\operatorname{double}$ maps stabilizers of $B$ to stabilizers of $D$ and logical operators of $B$ to logical operators of $D$. However, neither function preserves \emph{nontriviality}: a nontrivial logical operator of $B$ may double to a stabilizer of $D$, and a nontrivial logical operator of $D$ may fold to a stabilizer of $B$. The hypotheses of the doubling rule constrain these two events. This motivates the following definition:

\begin{definition}[seam logical]\label{def:seam}
Let $D$ be a double cover of $B$. We call a nontrivial logical operator $u \in C(G_B) \setminus G_B$ of the base a \emph{seam logical} if $\operatorname{double}(u)$ is a stabilizer of $D$, i.e., $\operatorname{double}(u) \in G_D$.
\end{definition}

For a seam logical $u$, applying $u$ to both sheets is a stabilizer of $D$, so a copy of $u$ on sheet $0$ differs from a copy of $u$ on sheet $1$ only by a stabilizer. Thus, $u$ can be slid across the seam from one copy of $B$ to the other without changing the logical operator it represents.

The distance of a double cover is governed by what its nontrivial logical operators fold to. Two of the possibilities give quantities of their own.

\begin{definition}[fold floor, stabilizer-fold floor]\label{def:floors}
Let $D$ be a double cover of $B$. The \emph{fold floor} $\phi(D)$ is the least weight of a nontrivial logical operator $b$ of $B$ such that $b = \operatorname{fold}(v)$ for some logical operator $v$ of $D$. The \emph{stabilizer-fold floor} $\rho(D)$ is the least weight of a nontrivial logical operator $v$ of $D$ such that $\operatorname{fold}(v) \in G_B \setminus \{I\}$. Each floor is $\infty$ if no such operator exists.
\end{definition}

\begin{theorem}[distance of a double cover]\label{thm:cover-bound}
Let $D$ be a double cover of $B$ and let $d := d(B)$. Then
\[
  d(D) \;\geq\; d' \;:=\; \min\{\phi(D),\; 2d,\; \rho(D)\}.
\]
If moreover some nontrivial logical operator of $B$ of weight $d$ is not a seam logical, then $d(D) \leq 2d$.
\end{theorem}

\begin{proof}
Let $v$ be a nontrivial logical operator of $D$ and let $b := \operatorname{fold}(v)$, a logical operator of $B$. \Cref{tab:fold-cases} lists the cases. If $b = I$, the two halves of $v$ agree, so $v = \operatorname{double}(u)$ for the operator $u$ of $v$ on sheet $0$. By item~5 above, $u$ commutes with $\operatorname{fold}(g)$ for every $g \in G_D$, and these folds are all of $G_B$, so $u$ is a logical operator of $B$. It is not a stabilizer of $B$: if $u = \operatorname{fold}(g)$ with $g \in G_D$, then $v = \operatorname{double}(\operatorname{fold}(g)) = g \cdot g^\sigma \in G_D$ by item~2 and swap-closure, contradicting the nontriviality of $v$. Hence $\mathrm{wt}(v) = 2\,\mathrm{wt}(u) \geq 2d$. If $b$ is a nontrivial logical operator of $B$, then $\mathrm{wt}(v) \geq \mathrm{wt}(b) \geq \phi(D)$ by item~4 and the definition of $\phi$. If $b$ is a stabilizer of $B$ other than the identity, then $\mathrm{wt}(v) \geq \rho(D)$ by definition. For the last claim, let $u$ be a nontrivial logical operator of $B$ of weight $d$ that is not a seam logical. Then $\operatorname{double}(u)$ is a logical operator of $D$ by item~5, it is not a stabilizer of $D$ because $u$ is not a seam logical, and it has weight $2d$ by item~3.
\end{proof}

The stabilizer-fold floor involves only light stabilizers: a nontrivial logical operator of $D$ that folds to a stabilizer $b$ has weight at least $\mathrm{wt}(b)$, so $\rho(D) < 2d$ can only be caused by stabilizers of $B$ of weight below $2d$. The fold floor is a quantity of $B$ once the mirror symmetry of $D$ extends to its logical operators.

\begin{lemma}[fold floor and seam logicals]\label[lemma]{lem:fold-floor}
Let $D$ be a double cover of $B$ such that $v \cdot v^\sigma \in G_D$ for every logical operator $v$ of $D$. Then every nontrivial logical operator of $B$ that is the fold of a logical operator of $D$ is a seam logical, and hence $\phi(D)$ is at least the weight of the lightest seam logical.
\end{lemma}

\begin{proof}
Let $b = \operatorname{fold}(v)$ be a nontrivial logical operator of $B$ with $v$ a logical operator of $D$. By item~2, $\operatorname{double}(b) = \operatorname{double}(\operatorname{fold}(v)) = v \cdot v^\sigma \in G_D$, so $b$ is a seam logical.
\end{proof}

The doubling rule of \cref{sec:doubling} is the case of \cref{thm:cover-bound} in which every term of the bound is at least $2d$.

\begin{corollary}[the doubling rule]\label[corollary]{cor:doubling}
Let $D$ be a double cover of $B$, and let $d := d(B)$. If
\renewcommand{\labelenumi}{\upshape(\theenumi)}
\begin{enumerate}
  \item\label{hyp:swap} $v \cdot v^\sigma \in G_D$ for every logical operator $v$ of $D$, i.e., the swap fixes every logical operator of $D$ up to a stabilizer;
  \item\label{hyp:seam} every seam logical has weight at least $2d$; and
  \item\label{hyp:stab-fold} for every stabilizer $b \in G_B$ with $0 < \mathrm{wt}(b) < 2d$, every nontrivial logical operator of $D$ that folds to $b$ has weight at least $2d$,
\end{enumerate}
then $d(D) = 2d$.
\end{corollary}

\begin{proof}
By \cref{lem:fold-floor}, hypotheses~(\ref{hyp:swap}) and~(\ref{hyp:seam}) give $\phi(D) \geq 2d$. Hypothesis~(\ref{hyp:stab-fold}) gives $\rho(D) \geq 2d$, because a nontrivial logical operator of $D$ that folds to a stabilizer of weight at least $2d$ already has weight at least $2d$. \Cref{thm:cover-bound} now gives $d(D) \geq 2d$. For the upper bound, a nontrivial logical operator of $B$ of weight $d$ exists by the definition of $d$, and it is not a seam logical by hypothesis~(\ref{hyp:seam}), since $d < 2d$; so \cref{thm:cover-bound} also gives $d(D) \leq 2d$.
\end{proof}

 We make three remarks on the hypotheses of \cref{cor:doubling}. First, hypothesis~(\ref{hyp:stab-fold}) is also necessary for $d(D) = 2d$, since it only asks that certain nontrivial logical operators of $D$ have weight at least $2d$. Second, hypotheses~(\ref{hyp:swap}) and~(\ref{hyp:seam}) are sufficient but not necessary: they bound the fold floor, and a light seam logical need not be the fold of any equally light logical operator of $D$, in which case the distance of $D$ exceeds the fold floor. Third, the nontrivial logical operators of $D$ that fold to the identity are  the doubles of the nontrivial logical operators of $B$ that are not seam logicals, so the term $2d$ in \cref{thm:cover-bound} is attained exactly when some such operator has weight $d$, which hypothesis~(\ref{hyp:seam}) guarantees; without it, that term can be replaced by twice the weight of the lightest such operator.

\paragraph{The Rule as a Computation on the Base Code}
The hypotheses of \cref{cor:doubling} are stated in terms of the logical operators of $D$. Read literally, checking them means reasoning about operators on the $2n$ qubits of $D$. This is not necessary. A double cover is determined by data on the base code: the generators $b_1, \ldots, b_r$ of $G_B$, and the operators $a_1, \ldots, a_r$ that say how each generator is split across the two sheets, giving the new stabilizer generators of $G_D$, $s_i = (a_i, a_i \cdot b_i)$. We now rewrite each hypothesis as a statement about operators on the $n$ qubits of $B$.

\emph{Finding the seam logicals.} Recall that a seam logical is a nontrivial logical operator of $B$ whose double is a stabilizer of $D$. Seam logicals come from redundancy among the generators of $B$. Suppose some of the generators multiply to the identity, $\prod_{i \in R} b_i = I$; we call such a set $R$ a \emph{relation}. For example, in the toric code the product of all the $X$-type generators is the identity. Multiplying the lifts of the generators in $R$ gives a stabilizer of $D$ that folds to the identity, so it acts in the same way on both sheets:
\[
  \prod_{i \in R} s_i \;=\; \operatorname{double}(\alpha_R), \qquad \text{where } \alpha_R := \prod_{i \in R} a_i .
\]
Conversely, every operator of $B$ whose double is a stabilizer of $D$ is a product of stabilizers of $B$ and operators $\alpha_R$. We call the group of all such products the \emph{seam group} and write $T$ for it. The seam logicals are the elements of $T$ that are not stabilizers of $B$. Computing $T$ needs only linear algebra, since the relations are the solutions of a linear system over $\mathbb{F}_2$.

\emph{Hypothesis~(\ref{hyp:swap}).} This hypothesis asks that swapping the sheets changes a logical operator $v$ of $D$ only by a stabilizer. Since $v \cdot v^\sigma = \operatorname{double}(\operatorname{fold}(v))$, it says that the fold of every logical operator of $D$ lies in the seam group $T$. The operators of $B$ that arise as such folds are exactly those that commute with every element of $T$. (An operator $p$ is the fold of a logical operator of $D$ when one can solve for the part of that operator on sheet~$1$, and this linear system has a solution if and only if $p$ commutes with $G_B$ and with every $\alpha_R$.) So the hypothesis holds if and only if every operator of $B$ that commutes with $T$ lies in $T$, that is, $C(T) \subseteq T$. This is again linear algebra and involves no weights.

\emph{Hypothesis~(\ref{hyp:seam}).} This hypothesis asks that every seam logical has weight at least $2d$. With the seam group in hand, it is a minimum-weight search over the elements of $T$ that are not stabilizers of $B$, all of which are operators of $B$.

\emph{Hypothesis~(\ref{hyp:stab-fold}).} This is the hypothesis that most visibly concerns $D$: it bounds the weight of the nontrivial logical operators of $D$ that fold to a light stabilizer $b$ of $B$. The key observation is that all of these operators have the same shape. Fix one stabilizer of $D$ that folds to $b$, say $s_b = (a_b,\ a_b \cdot b)$, obtained by multiplying the lifts of generators whose product is $b$. If $v$ is a nontrivial logical operator of $D$ that folds to $b$, then $v \cdot s_b$ is a nontrivial logical operator of $D$ that folds to the identity. By the first case in the proof of \cref{thm:cover-bound}, it is the double of a nontrivial logical operator $w$ of $B$, and $w$ is not a seam logical. Hence
\[
  v \;=\; s_b \cdot \operatorname{double}(w) \;=\; (a_b \cdot w,\ a_b \cdot b \cdot w),
\]
and every such $w$ gives an operator $v$ of this kind. So instead of searching over the logical operators of $D$, we search over the nontrivial logical operators $w$ of $B$ that are not seam logicals, and ask that $\mathrm{wt}(a_b \cdot w) + \mathrm{wt}(a_b \cdot b \cdot w) \geq 2d$.

This search is also shallow. On a qubit in the support of $b$, the two sheets of $v$ differ, so $v$ acts on at least one of the two copies; these qubits already contribute $\mathrm{wt}(b)$ to the weight of $v$. On a qubit outside the support of $b$, the two sheets of $v$ agree, so the qubit contributes $2$ if $a_b \cdot w$ acts on it and $0$ otherwise. It therefore suffices that $a_b \cdot w$ acts on at least $d - \mathrm{wt}(b)/2$ qubits outside the support of $b$.

\paragraph{Utility of the Proof Rule} After this translation, hypothesis~(\ref{hyp:swap}) is a rank computation and hypotheses~(\ref{hyp:seam}) and~(\ref{hyp:stab-fold}) are still minimum-weight problems, of the same kind as computing $d(B)$. Such problems are hard in general but they are posed on the $n$ qubits of $B$ rather than the $2n$ qubits of $D$. For the gross code, \f discharges the rule entirely on the $72$-qubit base. Hypothesis~(\ref{hyp:swap}) follows from two polynomial identities, and hypothesis~(\ref{hyp:seam}) from one kernel-checked certificate. This certificate also lists the $252$ light $X$-type stabilizers of the base, and a counting argument discharges hypothesis~(\ref{hyp:stab-fold}) for each of them without any search.

\paragraph{The Number of Logical Qubits}
The rules of \cref{fig:rules-alt} write $k'$ for the number of logical qubits of a double cover $D$ of $B$. This is determined by $k$ and the seam group. We show that $k' = 2k - t$, where $t$ is the number of independent generators of $T$ beyond those of $G_B$, that is, the number of independent seam logicals. Indeed, folding maps $G_D$ onto $G_B$, and the stabilizers of $D$ that fold to the identity are the doubles of the elements of $T$, so $G_D$ has $(n - k) + (n - k + t)$ independent generators on $2n$ qubits. Hence $k'$ ranges from $2k$, for two disjoint copies of $B$, down to $0$. Hypothesis~(\ref{hyp:swap}) of \cref{cor:doubling} forces $k' \leq k$, because it places the folds of the logical operators of $D$, a group with $n + k' - k$ independent generators, inside $T$, which has $n + k - k'$. The gross code attains $k' = k$.

\begin{table}[t]
\small
\caption{Folding a nontrivial logical operator $v$ of a double cover $D$ of $B$, where $d = d(B)$. Each row is one term of the bound in \cref{thm:cover-bound}; the hypotheses of \cref{cor:doubling} raise every term to $2d$.}
\label{tab:fold-cases}
\begin{tabular}{@{}p{0.3\linewidth}p{0.64\linewidth}@{}}
  \toprule
  $v$ folds to \ldots & so $v$ has weight at least \ldots \\
  \midrule
  the identity & $2d$: $v = (u, u)$ is the double of a nontrivial logical operator $u$ of $B$, so its weight is $2\,\mathrm{wt}(u) \geq 2d$. \\
  a nontrivial logical operator & $\phi(D)$, the fold floor, since folding never increases weight. Under hypothesis~(\ref{hyp:swap}) the fold is a seam logical, and hypothesis~(\ref{hyp:seam}) makes its weight at least $2d$. \\
  a stabilizer of weight $\geq 2d$ & $2d$, since folding never increases weight. \\
  a stabilizer of weight $1$ to $2d - 1$ & $\rho(D)$, the stabilizer-fold floor, by definition. Hypothesis~(\ref{hyp:stab-fold}) makes it at least $2d$. \\
  \bottomrule
\end{tabular}
\end{table}

\section{Proof Sketch for the Distance of the $\nkd{360, 12, 24}$ \textsc{bb} Code}\label{app:bb360}

We sketch the \f proof that the $\nkd{360, 12, 24}$ \textsc{bb} code of \cref{sec:BB} has distance exactly $24$. The upper bound is an explicit logical operator of weight $24$. For the lower bound, a symmetry between the $X$-type and $Z$-type operators of a \textsc{bb} code and a parity argument reduce the problem to showing that every $X$-type logical operator of weight at most $22$ is a stabilizer.

The code, which we write $C_{360}$, is a double cover of a $180$-qubit \textsc{bb} code $C_{180}$, which is in turn a double cover of a $90$-qubit \textsc{bb} code $C_{90}$. The two intermediate codes are obtained from the polynomials of $C_{360}$ by reducing $y$ modulo $3$ and then $x$ modulo $15$:
\[
\begin{array}{@{}llll@{}}
  C_{360} & \mathbb{Z}_{30} \times \mathbb{Z}_{6} & x^9 + y + y^2,\;\; y^3 + x^{25} + x^{26} & \nkd{360, 12, 24} \\
  C_{180} & \mathbb{Z}_{30} \times \mathbb{Z}_{3} & x^9 + y + y^2,\;\; 1 + x^{25} + x^{26} & \nkd{180, 8, 12} \\
  C_{90}  & \mathbb{Z}_{15} \times \mathbb{Z}_{3} & x^9 + y + y^2,\;\; 1 + x^{10} + x^{11} & \nkd{90, 8, 8}
\end{array}
\]
Of the parameters of $C_{180}$ and $C_{90}$, the proof uses only the lower bounds $12$ and $8$ on their distances. The proof can be viewed as answering the following question: given an operator $b$ of $B$, which logical operators $v$ of $D$ of weight at most $w$ fold to $b$? It answers this question for $D = C_{360}$, $B = C_{180}$, and shows that every such $v$ is a stabilizer, following the case analysis of \cref{thm:cover-bound}.

If $b$ is the identity, then $v$ is the double of a logical operator of $C_{180}$. Every nontrivial logical operator of $C_{180}$ has weight at least $12$, which \f proves by the same argument one level down, starting from a direct enumeration on $C_{90}$. So $v$ would have weight at least $24$ if it were nontrivial.

The proof does not use the hypotheses of \cref{cor:doubling} and instead uses an enumeration-based approach. Since folding never increases weight, $b$ has weight at most $22$, and so does its fold in $C_{90}$. A census lists every operator of $C_{90}$ that can arise in this way, up to translation of the torus. It has $274{,}011$ representatives, which a counting argument shows is complete. Lifting each representative to $C_{180}$ gives a finite list of candidates for $b$, and for each candidate the proof shows that every $v$ is a stabilizer. These searches are shallow for the reason given in \cref{app:double-cover}: an operator that folds to $b$ has weight $\mathrm{wt}(b)$ plus twice the number of qubits outside the support of $b$ on which it acts. The census and the searches are computed by untrusted programs outside Lean, and their results are checked by Lean's kernel.

\section{Lean Declarations of the Formalized Codes}\label{app:lean-names}

For each code of \cref{tab:case-studies}, \cref{tab:lean-names} names the \f declaration that packages the code with all three of its parameters.

\begin{table}[h]
\caption{The Lean declaration of type \lstinline|Code[[n, k, d]]| for each code of \cref{tab:case-studies}. Names are relative to the namespace \lstinline|Quantum|.}
\label{tab:lean-names}
\footnotesize
\begin{tabular}{@{}ll@{}}
  \toprule
  Code (family) & Lean declaration \\
  \midrule
  Gross code (\textsc{bb} family) & \lstinline|Stabilizer.Homological.BB.grossStabilizerCodeWithDistance| \\
  \addlinespace[0.5em]
  $360$-qubit \textsc{bb} code & \lstinline|Stabilizer.Homological.BB.Bravyi360.stabilizerCodeWithDistance| \\
  \addlinespace[0.5em]
  Mitten code & \lstinline|Stabilizer.Homological.LP.M150.mitten150StabilizerCodeWithDistance| \\
  \addlinespace[0.5em]
  Rotated surface code family & \lstinline|StabilizerGroup.RotatedSurfaceCodeN.rotatedSurfaceStabilizerCodeWithDistance| \\
  \addlinespace[0.5em]
  Toric code family & \lstinline|StabilizerGroup.ToricCodeN.toricStabilizerCodeWithDistance| \\
  \addlinespace[0.5em]
  Triangular color code family & \lstinline|StabilizerGroup.ColorCode.stabilizerCodeWithDistance| \\
  \bottomrule
\end{tabular}
\end{table}

\end{document}

%% file: figures/toric-grid-ops.tikz
\begin{tikzpicture}[x=0.4cm, y=0.4cm]
  \def\L{5}\pgfmathtruncatemacro{\lm}{\L-1}
  \newcommand{\basegrid}{%
    \foreach \k in {0,...,\lm}
      \draw[black!70, line width=0.4pt] (\k,-0.5) -- (\k,\lm+0.5) (-0.5,\k) -- (\lm+0.5,\k);}
  \newcommand{\vertices}{%
    \foreach \a in {0,...,\lm} \foreach \b in {0,...,\lm}
      \fill[black!70] (\a,\b) circle[radius=0.9pt];}
  \fill[black!12] (0,1) rectangle (2,2);
  \basegrid
  \draw[black!75, line width=1.6pt, line cap=round] (0,1) rectangle (2,2);
  \draw[black!60, line width=0.9pt, dotted] (1,1) -- (1,2);
  \draw[ccred, line width=2pt, line cap=round] (-0.5,3) -- (\lm+0.5,3);
  \draw[ccgreen, line width=2pt, line cap=round] (3,-0.5) -- (3,\lm+0.5);
  \vertices
  \begin{scope}[shift={(7.5,0.3)}, lbl/.style={font=\small, anchor=west}]
    \filldraw[fill=black!12, draw=black!75, line width=1.6pt] (0,3.1) rectangle ++(1.2,0.8);
    \node[lbl] at (1.6,3.5) {stabilizer: boundary of two faces, weight $6$};
    \draw[ccred, line width=2pt, line cap=round] (0,2.0) -- (1.2,2.0);
    \node[lbl] at (1.6,2.0) {nontrivial logical: wraps horizontally, weight $L$};
    \draw[ccgreen, line width=2pt, line cap=round] (0,0.9) -- (1.2,0.9);
    \node[lbl] at (1.6,0.9) {nontrivial logical: wraps vertically, weight $L$};
  \end{scope}
\end{tikzpicture}

%% file: figures/bb-stencils.tex
\newcommand{\bbpanel}[9]{%
  \foreach \k in {0,...,5}
    \draw[black!70, line width=0.4pt] (\k,-0.5) -- (\k,5.5) (-0.5,\k) -- (5.5,\k);
  \begin{scope}[line width=1.1pt, line cap=round]
    \begin{scope}[shift={(#1,#2)}, ccred]
      \foreach \p in {#3} \draw[shift={\p}] (0,-0.5) -- (0,0.5);
      \foreach \p in {#4} \draw[shift={\p}] (-0.5,0) -- (0.5,0);
      \node[font=\scriptsize] at (0.5,0.5) {$X$};
    \end{scope}
    \begin{scope}[shift={(#5,#6)}, ccblue]
      \foreach \p in {#7} \draw[shift={\p}] (0,-0.5) -- (0,0.5);
      \foreach \p in {#8} \draw[shift={\p}] (-0.5,0) -- (0.5,0);
      \node[font=\scriptsize] at (0.5,0.5) {$Z$};
    \end{scope}
  \end{scope}
  \node[font=\scriptsize, anchor=north, align=center] at (2.5,-0.7) {#9};
}
\begin{tikzpicture}[x=0.26cm, y=0.26cm]
  \bbpanel{1}{1}{(0,0.5),(1,0.5)}{(0.5,0),(0.5,1)}
          {4}{3}{(0,0.5),(0,-0.5)}{(0.5,0),(-0.5,0)}
          {$A = 1 + x$\\$B = 1 + y$}
  \begin{scope}[shift={(8.5,0)}]
    \bbpanel{0}{0}{(3,0.5),(0,1.5),(0,2.5)}{(0.5,3),(1.5,0),(2.5,0)}
            {4}{4}{(0,-2.5),(-1,0.5),(-2,0.5)}{(-2.5,0),(0.5,-1),(0.5,-2)}
            {$A = x^3 + y + y^2$\\$B = y^3 + x + x^2$}
  \end{scope}
\end{tikzpicture}